\documentclass[prd,twocolumn,showpacs,aps,amssymb]{revtex4}

\usepackage{graphicx}
\usepackage{bm}
\begin{document}

\title{Small-scale clumps in the galactic halo and dark matter
annihilation}

\author{Veniamin Berezinsky}
 \email{berezinsky@lngs.infn.it}
 \affiliation{INFN, Laboratori Nazionali del Gran Sasso, I-67010
  Assergi (AQ), Italy\\
and  Institute for Nuclear Research of the RAS, Moscow, Russia}

 \author{Vyacheslav Dokuchaev}
 \email{dokuchaev@inr.npd.ac.ru}
 \author{Yury Eroshenko}
 \email{erosh@inr.npd.ac.ru}
 \affiliation{Institute for Nuclear Research of the RAS, Moscow, Russia}

\date{\today}

\begin{abstract}

Production of small-scale DM clumps is studied in the standard
cosmological scenario with an inflation-produced primeval fluctuation
spectrum. Special attention is given to the three following
problems: (i) The mass spectrum of small-scale clumps with $M
\lesssim 10^3 M_{\odot}$ is calculated with tidal destruction of
the clumps taken into account within the hierarchical model of
clump structure. Only 0.1 - 0.5\% of small clumps survive the
stage of tidal destruction in each logarithmic mass interval
$\Delta\ln M\sim1$. (ii) The mass distribution of clumps has a
cutoff at $M_{\rm min}$ due to diffusion of DM particles out of a
fluctuation and free streaming at later stage. $M_{\rm min}$ is a
model dependent quantity. In the case the neutralino, considered
as a pure bino, is a DM particle, $M_{\rm min} \sim 10^{-8}
M_{\odot}$. (iii) The evolution of density profile in a DM clump
does not result in the singularity because of formation of the
core under influence of tidal interaction. The radius of the core
is $R_c \sim 0.1 R$, where $R$ is radius of the clump. The
applications for annihilation of DM particles in the Galactic
halo are studied. The number density of clumps as a function of
their mass, radius and distance to the Galactic center is
presented. The enhancement of annihilation signal due to
clumpiness, valid for arbitrary DM particles, is calculated. In
spite of small survival probability, the global annihilation
signal in most cases is dominated by clumps.  For observationally
preferable value of index of primeval fluctuation spectrum $n_p
\approx 1$, the enhancement of annihilation signal is described
by factor 2 - 5 for different density profiles in a clump.

\end{abstract}

\maketitle

\section{Introduction}

Both analytic calculations \cite{silk93,ufn2} and numerical
simulations \cite{moore1,ghigna98,klypin99} predict existence of
dark matter clumps in the Galactic halo. The density profile in
these clumps according to analytic calculations
\cite{bert85,hoff85,ryden87,ufn1} and numerical simulations
\cite{nfw,ghigna98} is $\rho(r) \propto r^{-\beta}$. An average
density of the dark matter (DM) in a galactic halo itself also
exhibits a similar density profile (relative to a galactic
center) in the both approaches. The DM profiles in clusters of
galaxies is discussed in \cite{kelson} and in references therein.
In the analytic approach of Gurevich and Zybin (see review
\cite{ufn1} and references therein) the density profiles are
predicted to be universal, with $\beta \approx 1.7 - 1.9$ for
clumps, galaxies and two-point correlation functions of galaxies.
In the numerical simulations the density profiles can be
evaluated only for the relatively large scales due to the limited
mass resolution. The value of $\beta$ differs in different
simulations from $\beta=1.0$ \cite{nfw} to $\beta=1.5$
\cite{moore1} and may be non-universal for the objects of
different mass scales \cite{JingSuto}. An attempt of analytical
explanation of the results of numerical simulations has been
performed in \cite{syer98,dekel02}. The phase-space density
profiles of DM halos are investigated in \cite{taylor01}.

A central cusp in the Galactic halo and the smaller scale clumps
result in the enhancement of DM annihilation rate and thus in the
stronger signals in the form of gamma-rays, radio emission,
positrons and antiprotons. The gamma-ray and radio signal from
central cusp in the Galactic halo has been first discussed in
\cite{BGZ,BBM94}. Recently this problem was examined in
\cite{berg,gs2000,zhao,bertone,merritt}. The enhancement of DM
annihilation rate due to the clumpiness of DM halo was first
pointed out in \cite{silk93}. Neutralino annihilation in clumps
can result in the very large diffuse gamma-ray flux \cite{BBM97}
in the model of the clumpy DM by Gurevich et al. \cite{ufn2}.
Calculations of positron and antiproton production in the clumpy
DM halo have been performed e.~g. in \cite{berg} (see also
\cite{JuKa,torino1,torino2}). Recently the annihilation of DM in
the clumps has been studied in
\cite{ABO,olinto,silk02,berg98,berg2001,berg2002}. The
synchrotron flux from DM annihilation products in clumps in the
presence of the Galactic magnetic field is considered in
\cite{bot02}. Constraints on the DM clumpiness in the halos from
heating of the disk galaxies is examined in \cite{cl1,atb02}.

The main purpose of this work is evaluation of the enhancement of
the annihilation signal due to presence of the small clumps of DM
in the Galactic halo.

The small-scale self-gravitating Dark Matter Clumps, which will be
referred to as DMCs or simply as clumps, may be formed in the
early universe due to several mechanisms. These DMCs may be
formed (i) by the growth of the adiabatic or isothermal
fluctuations (originated at inflation) during the
matter-dominated epoch; or (ii) from the density fluctuations in
the models with topological defects (cosmic strings and domain
walls) \cite{silk93}; or (iii) during the radiation dominating era
from the nonlinear isothermal fluctuations (originated by phase
transitions in early Universe) \cite{kolb94} or from the large
amplitude adiabatic fluctuations \cite{union}.

In this paper we shall consider only the most conservative case of
adiabatic fluctuations which enter the non-linear stage of
evolution at the matter dominated epoch with the
inflation-induced initial power-law power spectrum.

Small-scale clumps are formed only if the fluctuation amplitudes
in the spectrum are large enough at the corresponding small
scales. The inflation models predict the power-law primeval
fluctuation spectrum. If the power-law index $n_p \geq 1$, DMCs
are formed in a wide range of scales. During the universe
expansion the small clumps are captured by the larger ones, and
the larger clumps consist of the smaller ones and of continuously
distributed DM. The convenient analytic formalism, which
describes statistically this hierarchical clustering, is the
Press-Schechter theory \cite{ps74} and its extensions, in
particular `excursion set' formalism developed by Bond et al.
\cite{bond91} (for the clear introduction see \cite{cole}).
However, this theory does not include the important process of
the tidal destruction of small clumps inside the bigger ones. We
take into account this process in Sec.~\ref{razrush} and obtain
the mass function for the small-scale DMCs in the Galactic halo.
In the case of the power-law spectrum only a small fraction of
the captured clumps survives, but even this small fraction is
enough to dominate the total annihilation rate in the Galactic
halo.

In the hierarchical theory of large-scale structure formation in
the Universe the first formed objects have some minimal mass
$M_{\text{min}}$. The value of this mass is determined by the
spectrum of initial fluctuations and by the properties of DM
particles \cite{fsfirst,ufn2}. This value is crucial for
calculation of the DM annihilation rate. The estimates of $M_{\rm
min}$ existing in the literature for neutralino DM are
substantially different, from $M_{\text{min}}\sim
10^{-12}M_{\odot}$ in \cite{gzv12} to
$M_{\text{min}}\sim(10^{-7}-10^{-6})M_{\odot}$ in \cite{bino}. In
Sec.~\ref{smmin} we present our calculations and discuss the
previous results.

The DM annihilation rate crucially depends on the density profile
$\rho(r)\propto r^{-\beta}$ of DM particles in a clump and on the
distance $R_c$ where the density growth is cut off.  This region
is called the core. The radius of the core has been estimated in
the literature in the different approximations.  The estimation,
$R_c/R\sim\delta_{\text{eq}}^3$, where $\delta_{\text{eq}}$ is the
density fluctuation amplitude at the end of radiation dominated
epoch, has been obtained in \cite{ufn1}.  It was found from
behavior of the damped mode of nonlinear fluctuations.   A black
hole or baryonic core in the center of the DMC can strongly
affect the density saturation at $r\to 0$ for a very massive DMC
\cite{ufn1,gs2000,zhao,bertone}. Calculations \cite{BGZ,BBM97} of
the inward flux of DM particles into the dense central region of
DMC also result, due to annihilation of DM particles, in a very
small radius of the central core $R_c$.  The above mentioned
process is essential for the formation of the DMC core only in
the case of almost perfectly spherically symmetrical clump.

We shall estimate the radius of the core imposed by tidal
interaction, which gives the largest $R_c$ among those known in
the literature. In the spherically symmetric self-gravitating
clump on the stage of its formation the non-dissipative DM
particles are moving nearly radially in the oscillation regime.
The presence of a non-spherical (tidal) external gravitational
field causes the deflection of particle trajectories in the clump
from the radial ones. This process prevents the development of the
central singularity in a clump and results in the core formation.

During the radiation dominated epoch the small fluctuations
$\delta=\delta\rho/\rho\ll 1$ grow very slowly. At the matter
dominated stage these fluctuations start to grow fast
\cite{svvbm} in the regime $\delta\propto t^{2/3}$. Fluctuations
get detached from the general cosmological expansion and start
contracting after reaching the nonlinear value, $\delta\geq 1$.
These nonlinear fluctuations form finally the DM clumps. The
analytic studies of nonlinear evolution of fluctuations have been
performed by many authors. One of the most detailed analytical
approach was developed by Gurevich and Zybin \cite{ufn1}. In this
formalism at the certain moment of gravitational contraction the
density singularity forms in the center of a nonlinear
fluctuation. From this singularity point the density caustics
(i.~e. the boundary of regions with a different number of
streams) expand outward. The secondary caustics appear inside the
primary ones and their number increases fast with time. This
multi-stream instability has been discovered and studied in
detail in \cite{ufn1}. It was demonstrated that the stationary
universal density profile $\rho(r)\propto r^{-\beta}$ with
$\beta=1.7- 1.9$ is formed as a result of streams mixing. The
maximum density of DM particles in a clump is reached at the
center.

In our consideration of the clump formation we shall follow for
convenience the theoretical scenario of Gurevich and Zybin
\cite{ufn1}. However, the effects of tidal interaction, which is
the main result of our work, are valid for a much broader class
of scenarios.

The processes described above are valid for all DM particles
which can be considered as non-dissipative. The signal production
depends on the annihilation cross-section and thus on the nature
of DM particles. However our strategy is to calculate the
enhancement of the signal due to the halo clumpiness in
comparison with an isotropic unclumped distribution of DM. As a
guide we shall take neutralino as DM particle, but essentially our
results for enhancement of the annihilation signal are relevant
for a wide class of other DM particles.

We perform our calculations for cosmological model with the
matter density $\Omega_{\mathrm{m}}=0.3$ and the cosmological
term $\Omega_{\Lambda} = 1-\Omega_{\mathrm{m}}\simeq0.7$. The
presence of $\Lambda$ term influences only the value of
$\rho_{\mathrm{\text{eq}}}$ and does not affect the formation of
low mass DMCs. This is because the $\Lambda$ term contributes
negligibly to the total cosmological density at time scales when
the low-mass DMCs formation occurs. We shall use index `eq' for
the values at the moment of equality (i.~e. transition from the
radiation dominated to matter dominated epoch). We shall use the
Hubble constant 70~km~s~$^{-1}$~Mpc~$^{-1}$.

\section{Enhancement of Annihilation Signal due to Clumps}
\label{ampl}

Let us consider a DM clump with the internal density profile
$\rho_{\rm int}(r)$ and a total mass $M=\int4\pi r^2\rho_{\rm
int}(r)dr$. An annihilation rate in a single clump is given by
\begin{equation}
\dot N_{\rm cl}\!=\!4\pi\int\limits_{0}^{\infty}\!r^2dr\rho_{\rm
int}^2(r) m_{\chi}^{-2}\langle\sigma_{\text{ann}} v\rangle\! =\!
\frac{3}{4\pi} \frac{\langle\sigma_{\text{ann}}
v\rangle}{m_{\chi}^2} \frac{M^2}{R^3}\,S, \label{separ}
\end{equation}
where $m_{\chi}$ is mass of a DM particle (being not necessarily
neutralino), $v$ is relative velocity of two DM particles at the
collision, $\sigma_{\rm ann}$ is annihilation cross-section and
$R$ is virial radius of a clump. Function $S$ is determined by
Eq.~(\ref{separ}) and depends on DM distribution in a clump, in
particular $S=1$ for the simplest case of an uniform clump, when
$\rho_{\rm int}(r)=const$ at $r\leq R$ and $\rho_{\rm int}(r)=0$
at $r>R$.

An expansion of $\langle\sigma_{\rm ann} v\rangle$ over the
relative velocity $v$ of two DM particles has a form
\begin{equation}
\langle\sigma_{\rm ann}v\rangle= a+bv^2+cv^4+...
\label{cross-sect}
\end{equation}
where $a$ has a contribution of s-wave amplitude only, and $b$
--- from
both s- and p- waves. Since $v$ is very small, $\langle\sigma_{\rm
ann}v\rangle$ can be put out of the integral in Eq.~(\ref{separ}).

We shall use the following parametrization of the density profile
in a clump
\begin{equation}
\rho_{\rm int}(r)=\left\{
\begin{array}{lr}
\rho_c, & r<R_c; \\
\displaystyle{\rho_c\left(\frac{r}{R_c}\right)^{-\beta}}, & R_c<r<R;\\
 0, & r>R.
 \label{rho}
\end{array}
\right.
\end{equation}
Using $\rho_{\rm int}(r)$ from Eq.~(\ref{rho}) it is easy to
calculate $S$ from Eq.~(\ref{separ}) as
\begin{equation}
S(x_c,\beta)\!=\! \frac{(3-\beta)^2}{3(2\beta-3)}
\left(\frac{2\beta}{3}x_c^{3-2\beta}-\!1\!\!\right)\!\!
\left(\!1\!-\!\frac{\beta}{3}x_c^{3-\beta}\right)^{-2}\!\!\!\!\!,
\label{sbig}
\end{equation}
where $x_c=R_c/R$. Another approach to the parametrization of
clump structure has been used in \cite{silk02}.

There is distribution of clumps in the Galactic halo at least
over three parameters, mass $M$, radius $R$, and distance from
the Galactic Center $l$ : $n_{\rm cl}(M,R,l)$. This distribution
can also depend on the parameters which describe the internal
structure of the clumps, $\beta$ and $x_c$, from Eq.~(\ref{rho}).
We shall discuss this dependence in Sec.~\ref{strukt}. In
particular it will be demonstrated that $x_c=x_c(M,R)$, while
$\beta$ is the universal constant. Thus the differential number
density of DMCs in the halo can be written as:
\begin{equation}
dN_{\text{cl}}=n_{\text{cl}}(l,M,R)d^3ldMdR, \label{nclfirst}
\end{equation}
The observed signal at the position of the Earth from DM particle
annihilation in the clumps is proportional to the quantity
\begin{eqnarray}
I_{\rm cl}\!&=&\!\frac{1}{4\pi}
\int\limits_{0}^{\pi}\!d\zeta\sin\zeta\!\!\!
\int\limits_{0}^{r_{\text{max}}(\zeta)}\frac{2\pi r^2dr}{r^2}\!\!
\int\limits_{M_{\text{min}}}^{M_{\text{max}}}\!\!\!dM \!\!\!
\int\limits_{R_{\text{min}}}^{R_{\text{max}}}\!\!\! dR \nonumber \\
&& \times\,\, n_{\text{cl}}(l(\zeta,r),M,R) \dot N_{\rm cl}(M,R),
\label{ihal}
\end{eqnarray}
where $r$ is distance from the Sun (Earth) to a clump and $\zeta$
is angle between the line of observation and the direction to the
Galactic center. The distance $l$ between a clump and the
Galactic center can be given in terms of $r$, $r_{\odot}$
(distance from the Sun to the Galactic center) and $\zeta$ as
$l(\zeta,r) = (r^2+r_{\odot}^2-2rr_{\odot}\cos\zeta)^{1/2}$, and
the maximum distance from the Sun to the outer halo border in the
direction of $\zeta$, $r_{\text{max}}(\zeta) =
(R_H^2-r_{\odot}^2\sin^2\zeta)^{1/2}$, where $R_H \sim 100$~kpc
is the Galactic halo virial radius and $r_{\odot}=8.5$~kpc is the
distance from the Sun to the Galactic center.

Additional annihilation signal is given by unclumpy DM in the halo
with homogeneous ({\em i.e.} smoothly spread) density $\rho_{\rm
DM}(l)$, where $l$ is a distance to the Galactic Center.
\begin{equation}
I_{\text{hom}}=\frac{\langle\sigma_{\rm ann}v\rangle}{2}
\int\limits_{0}^{\pi}\!
d\zeta\sin\zeta\!\!\!\int\limits_{0}^{r_{\text{max}}(\zeta)}
\!\!\!\!dr\rho_{\text{DM}}^2(l(\zeta,r))/m_{\chi}^2. \label{hom}
\end{equation}
The {\em enhancement} $\eta$ of the signal due to a presence of
clumps is given by
\begin{equation}
\eta=\frac{I_{\rm cl}+I_{\text{hom}}}{I_{\text{hom}}} \label{eta}
\end{equation}

This quantity describes the global enhancement of the
annihilation signal observed at the Earth ({\em e.~g.} the flux
of radio, gamma, and neutrino radiations) as compared with usual
calculations from annihilation of unclumpy DM. From
Eqs.~(\ref{eta}), (\ref{hom}) and (\ref{ihal}) one can see that
enhancement $\eta$ does not depend on the properties of DM
particles, in particular on the annihilation cross-section, and
is fully determined by the parameters of DM clumpiness. The
further exact calculations in this paper will be performed using
Eqs.~(\ref{ihal})--(\ref{eta}), but now we shall turn to the
approximate expression for $\eta$.

We shall accept now the simplifying assumptions. We assume that
space density of clumps in the halo, $n_{\rm cl}(l)$ is
proportional to the unclumpy DM density, $\rho_{\rm DM}(l)$:
$n_{\text{cl}}(l)=\xi\rho_{\rm DM}(l)/M$ with $\xi \ll 1$. This
assumption holds with a good accuracy for the small-scale clumps.
In contrast, the distribution of large-scale clumps obtained in
the numerical simulations \cite{moore99} is rather different from
the density distribution of the small clumps, especially in a
central part of the halo because of the tidal disruption of
clumps there. However, the clump signal is determined mostly by
clumps of the minimal mass. We neglect here the distribution of
clumps over $M$ and $R$. Instead we shall use a mean density of
DM particles inside a clump $\bar\rho_{\text{int}}=3M/4\pi R^3$.
Finally, we shall introduce for convenience the effective density
of DM particles in the halo defined as
\begin{equation}
\tilde\rho_{\rm DM}\equiv \frac{\displaystyle
\int\limits_{0}^{\pi}d\zeta\sin\zeta
\int\limits_{0}^{r_{\text{max}}(\zeta)}dr \rho_{\rm
DM}^2(l(\zeta,r)) } {\displaystyle
\int\limits_{0}^{\pi}d\zeta\sin\zeta
\int\limits_{0}^{r_{\text{max}}(\zeta)}dr \rho_{\rm
DM}(l(\zeta,r)) }. \label{rhoeff}
\end{equation}
As a result, we obtain for an enhancement factor the convenient
{\em estimate}
\begin{equation}
\eta \approx 1+\xi S(x_c,\beta)\frac{\bar\rho_{\text{int}}}
{\tilde\rho_{\rm DM}}, \label{etasimp}
\end{equation}
where $\xi$ is a fraction of DM in the form of clumps (see above)
and $S(x_c,\beta)$ is given by Eq.~(\ref{sbig}). For typical
parameters (see details in the following sections) $n_p \simeq 1$,
$\beta\simeq 1.8$, $x_c\simeq0.05$, $S(x_c,\beta)\simeq5$,
$\tilde\rho_H \sim
\rho_{\text{DM}}(r_{\odot})\sim0.3$~GeV~cm$^{-3}$,
$\bar\rho_{\text{int}}\sim2\times10^{-22}$~g~cm$^{-3}$,
$\xi\sim0.001$, the numerical estimate $\eta\sim 3$ follows from
Eq.~(\ref{etasimp}).

\section{Clumps of Minimal Masses}
\label{smmin}

The number of clumps in the halo increases at small clump masses
$M$, and the signal from clumps $I_{\rm cl}$ crucially depends on
$M_{\rm min}$ in the clump distribution as Eq.~(\ref{ihal})
shows. The value of  $M_{\rm min}$ is determined by a leakage of
DM particles from the overdense fluctuations in the early
universe. We shall describe first this process qualitatively and
then present numerical calculations.

CDM particles at high temperature $T>T_f \sim 0.05 m_{\chi}$ are
in the thermodynamical (chemical) equilibrium with cosmic plasma,
when their number density is determined by temperature. After
freezing at $t>t_f$ and $T<T_f$, the DM particles remain for some
time in {\em kinetic} equilibrium with plasma, when the
temperature of CDM particles $T_{\chi}$ is equal to temperature
of plasma $T$, but number density $n_{\chi}$ is determined by
freezing concentration and expansion of the universe. At this
stage the CDM particles are not perfectly coupled to the cosmic
plasma. Collisions between a CDM particle and fast particles of
ambient plasma result in exchange of momenta and a CDM particle
diffuses in the space. Due to diffusion the DM particles leak
from the small-scale fluctuations and thus their distribution
obtain a cutoff at the minimal mass $M_D$.

When the energy relaxation time for DM particles
$\tau_{\text{rel}}$ becomes larger than the Hubble time
$H^{-1}(t)$, the  DM particles get out of the kinetic
equilibrium. This conditions determines the time of kinetic
decoupling $t_d$. At $t \geq t_d$ the CDM matter particles are
moving in the free streaming regime and all fluctuations on the
scale of
\begin{equation}
\lambda_{\rm fs}=a(t_0)\int_{t_d}^{t_0}\frac{v(t')dt'}{a(t')}
\label{fs}
\end{equation}
and smaller are washed away [here $a(t)$ is the scaling factor of
expanding universe and $v(t)$ is velocity of a DM particle at
epoch $t$]. The corresponding minimal mass at epoch $t_0$,
\begin{equation}
M_{\rm fs}=\frac{4\pi}{3}\rho_{\chi}(t_0)\lambda_{\rm fs}^3,
\label{Mfs}
\end{equation}
is much larger than $M_D$. Numerical calculations below (for
neutralino) show that $M_D$ is close to $M_{\rm min}$ from
\cite{gzv12} and $M_{\rm fs}$ to $M_{\rm min}$ from \cite{bino}.

The calculation of the minimal mass $M_{\rm min}$ in the mass
spectrum of fluctuations is obviously model dependent. As the DM
particle we shall consider the neutralino $\chi$, for which we
take the pure bino state ($\chi=\tilde{B}$). As calculations
below show the temperature of kinetic decoupling for a reasonable
range of parameters is $T_d \sim 100 $~MeV, and thus we can
consider cosmic plasma consisting of relativistic electrons,
positrons, neutrinos and photons in thermal equilibrium.

The cross-sections for scattering of bino off left (right)
electron and left neutrino are given in the Appendix A. A
cross-section for $\nu\chi$ scattering is given by
Eq.~(\ref{crosss1}) and for $e\chi$ scattering it is 17 times
larger, if to assume equal masses of selectrons and sneutrinos
(we shall use $\tilde{m}$ for the both left and right selectron
and sneutrino masses, and $\tilde{M}^2=\tilde{m}^2-m_{\chi}^2$).

First of all we shall calculate the moment $t_d$ and temperature
$T_d$ of kinetic decoupling of neutralino, using condition
\begin{equation}
\frac{1}{\tau_{\text{rel}}}\simeq H(t), \label{tauH}
\end{equation}
where $H(t)=1/(2t)$ is the Hubble constant and
$\tau_{\text{rel}}(T)$ is the energy relaxation time for
neutralino at temperature of electron-neutrino gas $T$. The
relaxation time $\tau_{\text{rel}}$ is determined by collisions
of neutralino with fermions $\nu_L,~~e_L$ and $e_R$. Neutralino
can be considered as particle at rest because the rest system
coincides with the center-of-mass system with the accuracy of
order $\sqrt{T/m_{\chi}}$. Let $\delta p$ is the neutralino
momentum obtained in one scattering:
\begin{equation}
\label{dp2} (\delta p)^2=2\omega^2(1-\cos\theta),
\end{equation}
where $\omega$ and $\theta$ is neutrino energy and scattering
angle, respectively.

Let us introduce the number density of relativistic fermions with
one polarization and with energy $\omega$:
\begin{equation}
n_0(\omega)=\frac{1}{2\pi^2}\frac{\omega^2}{e^{\omega/T}+1},
\end{equation}
Then for the energy relaxation time $\tau_{\text{rel}}$ we have
\begin{equation}
\frac{1}{\tau_{\text{rel}}}\!=\!\frac{1}{E_k}\frac{d E_k}{dt}\!=
\!\frac{40}{2E_km_{\chi}}\! \int \!\!d\Omega\int\!\! d\omega\,
n_0(\omega) \!\! \left( \frac{d\sigma_{\text{el}}}{d\Omega}
\right)_{f_L\chi}\!\!\!\! (\delta p)^2, \label{tau}
\end{equation}
where $E_k\simeq(3/2)T$ is a mean kinetic energy of neutralino,
and $(d\sigma_{\text{el}}/d\Omega)_{f_L\chi}$ is given by
Eq.~(\ref{crosss1}). The number 40 in Eq.~(\ref{tau}) is obtained
by counting of degrees of freedom: three neutrinos and
antineutrinos (or $\nu_L^c$ in case of Majorana neutrinos) give
6, $e_L$ and $e_L^c$ give 2 and two right (singlet) states for
electron and positron gives 34, because their cross-sections are
17 times larger.

After integration in Eq.~(\ref{tau}) we obtain
\begin{equation}
\frac{1}{\tau_{\text{rel}}}=\frac{40\Gamma(7)\alpha_{\text{e.m.}}^2}
{9\pi\cos^4\theta_{\text{W}}} \frac{T^6}{\tilde M^4m_{\chi}}.
\label{taurel}
\end{equation}
Using Eq.~(\ref{tauH}) and connection between age and temperature
of the universe
\begin{equation}
t=\frac{2.42}{\sqrt{g_*}}\left(\frac{T} {1\mbox{
MeV}}\right)^{-2}\mbox{
 s}, \label{ttime}
\end{equation}
where $g_*$ is number of degrees of freedom, we obtain
numerically:
\begin{equation}
t_d=3.5\!\times\!10^{-5}\!\left(\frac{m_{\chi}}{100\mbox{
GeV}}\right)^{-1/2} \!\left(\frac{\tilde M}{1\mbox{
TeV}}\right)^{-2}
\!\!\left(\frac{g_*}{10}\right)^{-3/4}\!\!\!\mbox{s}, \label{tsmd}
\end{equation}
and
\begin{equation}
T_d=150\left(\frac{m_{\chi}}{100\mbox{ GeV}}\right)^{1/4}
\left(\frac{\tilde M}{1\mbox{ TeV}}\right)
\left(\frac{g_*}{10}\right)^{1/8}\mbox{ MeV}. \label{tbigd}
\end{equation}
We shall present in this section the calculations made in
physically transparent way, considering the diffusion leaking of
neutralinos at the stage of kinetic equilibrium and free
streaming when neutralinos get out of kinetic equilibrium. In
Appendix~B we shall study both stages together in the formalism
of kinetic equation, as it has been done in \cite{gzv12}. Though
our methods are not identical, their comparison implies that the
absence of free streaming is responsible for the contradiction
with different $M_{\rm min}$ discussed above. The independent
approach in Appendix~B confirms the results obtained below.

\subsection{Diffusion cutoff of the mass spectrum}
\label{M_D}

We can come over now to the calculation of $M_D$, the minimal
mass in the fluctuation spectrum caused by diffusion of
neutralinos out of an overdense fluctuation. We calculate the
diffusion coefficient using the method given in \cite{fk} (\S12).
Consider a neutralino moving with a nonrelativistic velocity
$\vec v$. In the rest system of this particle the momentum
distribution of relativistic fermions is anisotropic:
\begin{equation}
n(\vec p)\,d^3p=\frac{d\Omega_{\alpha}\,p^2\,dp}{(2\pi)^3}
\frac{1}{e^{p(1+v\cos\alpha)/T}+1},
\end{equation}
where $\alpha$ is the angle between the directions of $\vec v$ and
momentum of incoming fermion.

The momentum transfer in a single scattering equals to $\vec
p(1-\cos\theta)$ after averaging over the azimuthal angles.

A corresponding friction force experienced by the neutralino is
\begin{eqnarray}
\vec f_r& \!=\! & 40 \int\! d\Omega_{\theta}\int\! d^3p\, n(\vec
p)
\left(\frac{d\sigma_{\text{el}}}{d\Omega_{\theta}}\right)_{f_l\chi}
\vec p\,(1-\cos\theta) \nonumber \\
& \!=\! &-B^{-1}\vec v,
\end{eqnarray}
where $B$ is a particle mobility and factor 40 takes into account
scattering on all fermions as in Eq.~(\ref{taurel}). Then the
diffusion coefficient is
\begin{equation}
D=BT=\frac{3\pi\cos^4\theta_{\text{W}}\tilde M^4}{40\Gamma(6)
\alpha_{\text{e.m.}}^2T^5}. \label{ddd}
\end{equation}
Diffusion equation in the comoving system has a form
\begin{equation}
\frac{\partial\delta(\vec x,t)}{\partial t}=
\frac{D(t)}{a^2(t)}\Delta_{\vec x}\delta(\vec x,t). \label{diff}
\end{equation}
Diffusion coefficient $D(t)$ is time-dependent because of $T(t)$.
Solution of Eq.~(\ref{diff}) for the Fourier component is
\begin{equation}
\delta_{\vec k}(t)= \delta_{\vec k}(t_f)\exp\left\{
-k^2Cg_*^{5/4}\tilde M^4 \left(t^{5/2}-t_f^{5/2}\right) \right\},
\label{dkdif}
\end{equation}
where $C=const$. The factor
$Cg_*^{5/4}\tilde{M}^4(t^{5/2}-t_f^{5/2})$ in front of $k^2$ in
Eq.~(\ref{dkdif}) is the diffusion length squared
$\lambda_{\text{D}}^2(t)/a^2(t)$ in the comoving coordinates.
This value determines the minimal mass in the density
perturbation spectrum due to diffusion of neutralinos from a
fluctuation:
\begin{eqnarray}
M_{\text{D}}\!&=&\!\frac{4\pi}{3}\rho_{\chi}(t_d)\lambda_{\text{D}}^3(t_d)
=4.3\times\!10^{-13} \!\left( \frac{m_{\chi}}{100\mbox{
GeV}}\right)^{-15/8} \nonumber
\\
&\times&\!\! \!\!\left(\frac{\tilde M}{1\mbox{ TeV}}\right)^{-3/2}
\left(\frac{g_*}{10}\right)^{-15/16} M_{\odot}. \label{mdif}
\end{eqnarray}
The functional dependence of Eq.~(\ref{dkdif}) and numerical
value of (\ref{mdif}) obtained in diffusion approximation
coincide with the corresponding results obtained by different
method in \cite{gzv12}.

\subsection{Free streaming cutoff of the mass spectrum}
\label{FS}

We shall consider now the free streaming cutoff of the mass
spectrum qualitatively described in the beginning of this
section. We have given there an estimate of the minimal mass due
to free streaming. In the accurate calculations below we shall
take into account the angular and velocity distribution of
leaking neutralinos, and exact dependence of $a(t)$ at age $\sim
t_{\rm eq}$, which affect the value of $M_{\rm fs}$.

After the moment of kinetic decoupling $t_d$, neutralinos move
freely in the expanding universe background, $a(t)d\vec x=\vec
v(t)dt$, where $\vec x$ is comoving particle coordinates.
Coordinates $\vec x$ at some moment $t$ are determined by initial
coordinates $\vec q$ and velocity $\vec v_d$ at the moment of
kinetic decoupling $t_d$:
\begin{equation}
\vec x=\vec f(\vec q,\vec v_d,t)=\vec q+ \int
\limits_{t_d}^{t}\frac{\vec v(t')\,dt'}{a(t')} = \vec q+g(t)\vec
v_d, \label{deff}
\end{equation}
where
\begin{equation}
g(t)=a(t_d)\int \limits_{t_d}^{t}\frac{dt'}{a^2(t')}, \label{defg}
\end{equation}
$\vec v(t)=\vec v_da(t_d)/a(t)$ for nonrelativistic particle. Now
for the neutralino number density at the point $\vec x$ we have
\begin{eqnarray} \label{lagrn}
 n(\vec x,t)&=& \!\!\int\!\! d^3v_d\,\phi(\vec v_d) \sum \limits_{\vec
q_*} n(\vec q_*,t_d)\left|\frac{D\vec f}{D\vec q}~\right|_{\vec q
= \vec q_*}
\\
&=& \!\!\int \!\! d^3v_d\,\phi(\vec v_d)\!\int \!d^3 q~n(\vec
q,t_d)\delta^{(3)}(\vec x-\vec f(\vec q,\vec v_d,t)), \nonumber
\end{eqnarray}
where $\delta^{(3)}$ is the Dirac delta-function, $D\vec f/D\vec
q\,$ is the Jacobian and $\phi(\vec v_d)$ is neutralino velocity
distribution function at the moment $t_d$. Summation in
(\ref{lagrn}) goes over all roots $\vec q_*$ of the equation
$\vec x=\vec f(\vec q,\vec v_d,t)$ from (\ref{deff}). This sum in
fact has only one term because the function $f(\vec q,\vec
v_d,t)$ in (\ref{deff}) is a single-valued one.

From (\ref{deff}) and (\ref{lagrn}) we find the Fourier component
\begin{equation}
n_{\vec k}(t)=n_{\vec k}(t_d) \int d^3v_d\,\phi(\vec v_d)
e^{-i\vec k\vec v_dg(t)}. \label{promnk}
\end{equation}
Assuming velocity distribution at the moment of decoupling
$\phi(\vec v_d)$ to be Maxwellian
\begin{equation}
\phi(\vec v_d)=\left(\frac{m_{\chi}}{2\pi T_d}\right)^{3/2}
\exp\left\{-\frac{m_{\chi}v_d^2}{2T_d}\right\},
\end{equation}
we obtain
\begin{equation}
n_{\vec k}(t) = n_{\vec k}(t_d)
e^{-\frac{1}{2}k^2g^2(t)\frac{T_d}{m_{\chi}}}, \label{nk}
\end{equation}
i.e. up to the moment $t$ all perturbations are washed out by free
streaming inside the physical length-scale
\begin{equation}
\lambda_{\text{fs}}(t)=a(t)g(t)\left(\frac{T_d}{m_{\chi}}\right)^{1/2}.
\label{lambdafs}
\end{equation}
This length-scale corresponds to the clump of the minimal mass
\begin{equation}
M_{\text{fs}}(t)=\frac{4\pi}{3}
\rho_{\chi}(t)\lambda_{\text{fs}}^3(t), \label{mfs}
\end{equation}
where $\rho_{\chi}(t)=\rho_{\text{eq}}a_{\text{eq}}^3/a^3(t)$.
The time dependence of $M_{\text{fs}}(t)$ is regulated by $a(t)$.
At the radiation dominated epoch, $M_{\text{fs}}(t)$ grows
logarithmically with time. This growth is saturated at the matter
dominated epoch. The resulting $M_{\rm min}$ at $t_0$ can be
easily calculated using $a(t)$ as solution of the Friedman
equation:
\begin{equation}
\begin{array}{l}
\displaystyle{ a(\eta)=a_{\text{eq}}\left[ 2\frac{\eta}{\eta_*}
+\left(\frac{\eta}{\eta_*}\right)^2 \right]},
\\
\\
\displaystyle{ t = a_{\text{eq}}\eta_*
\left[\left(\frac{\eta}{\eta_*}\right)^2 +
\frac{1}{3}\left(\frac{\eta}{\eta_*}\right)^3 \right]}.
\end{array}
\label{frid}
\end{equation}
In these equations $\eta_*^{-2}=2\pi
G\rho_{\text{eq}}a_{\text{eq}}^2/3$, $a_{\text{eq}}$ is the value
of scaling factor at the moment $t_{\text{eq}}$,
\begin{equation}
\rho_{\text{eq}}\!=\!\rho_0(1+z_{\text{eq}})^3\!=\!1.1\times10^{-19}
\left(\frac{h}{0.7}\right)^{8}\!\!\left(\frac{\Omega_m}{0.3}\right)^{4}
\!\!\mbox{g cm}^{-3}, \label{kritden}
\end{equation}
$1+z_{\text{eq}}= 2.35\times10^4\Omega_mh^2$ and
$\rho_0=1.9\times10^{-29}\Omega_mh^2\mbox{ g cm}^{-3}$. The
presence of cosmological constant $\Lambda$ affects only the
value $\rho_{\text{eq}}$ and does not influence the evolution
$M_{\text{fs}}(t)$ because the contribution of $\Lambda$ to the
total cosmological density is negligible at small $t$. Putting
(\ref{frid}) into (\ref{defg}), we find after integration
\begin{equation}
M_{\text{min}}\!=\!\frac{\pi^{1/4}}{2^{19/4}3^{1/4}}
\frac{\rho_{\text{eq}}^{1/4}t_d^{3/2}}{G^{3/4}}
\left(\frac{T_d}{m_{\chi}}\right)^{3/2}\!\!\!\ln^3\left\{\frac{24}{\pi
G\rho_{\text{eq}}t_d^2} \right\}. \label{mminbukv}
\end{equation}
Using Eqs.~(\ref{tsmd}) and (\ref{tbigd}) we obtain numerically
\begin{eqnarray}
M_{\text{min}}&=&1.5\times10^{-8}\left(\frac{m_{\chi}}{100\mbox{
GeV}} \right)^{-15/8} \left(\frac{\tilde M}{1 \mbox{
TeV}}\right)^{-3/2}
\nonumber\\
&&\times\left(\frac{g_*}{10}\right)^{-15/16}
\left(\frac{\Lambda^*}{83}\right)^3 M_{\odot}, \label{mminnum}
\end{eqnarray}
where $\Lambda^*$ is the logarithm from Eq.~(\ref{mminbukv}).

Our calculations agree well with \cite{bino} as far as the most
important quantity $T_d$ is concerned (the scattering
cross-section is involved only there). The calculation of $M_{\rm
min}$ from $T_d$ in our case involves non-radial propagation of
neutralinos in a fluctuation and their distribution over
velocities, Eqs. (\ref{deff}) - (\ref{nk}). We include also the
accurate time dependence of the scaling factor $a(t)$. The
calculations in \cite{bino} follow the semi-quantitative scheme
described in the beginning of Sec. \ref{smmin}. At this stage of
calculations we have a difference described by factor 7.

{\em In conclusion}, in this section we have considered two
processes of washing out the cosmological density perturbations
in neutralino gas. The first process is the neutralino diffusion
due to scattering off neutrinos, electrons and positrons. This
process is effective until neutralinos are in the kinetic
equilibrium with the cosmological plasma. Up to the moment of
decoupling $t_d$ all perturbations with mass $M<M_{\text{D}}
\simeq 10^{-13}-10^{-12}M_{\odot}$ are washed out. The second
process is neutralino free steaming. Starting later, at $t>t_d$,
it washes out the larger perturbations with $M \leq M_{\rm fs}$
and determines $M_{\rm min}$ in the clump mass distribution at
present epoch, as given by Eq.~(\ref{mminnum}).

\section{Core of a Dark Matter Clump}
\label{strukt}

In this section we shall consider smearing of the singular density
profile in a clump due to tidal forces and calculate the radius
$R_c$ of the produced core.

Clumps, as well as galaxies, are originated near the maxima in
cosmological density perturbations $\delta(\vec r)=(\rho(\vec
r)-\bar\rho)/\bar\rho$.  At the matter dominated stage the
density perturbations grow as $\delta\propto t^{2/3}$.  In the
nonlinear regime, $\delta\gtrsim1$, the multi-flux instability
develops in a clump and singular density profile is formed
\cite{ufn1}.  If velocity field in the central part of the clump
is disturbed and becomes weakly nonradial, the flow is
overturned, singularity does not form and density profile is
smoothed.  In \cite{ufn1} the core radius is estimated as
$x_c\simeq\delta^3_{\text{eq}}\ll1$ from consideration of the
perturbation of the velocity field due to damped mode of the
cosmological density perturbations.  Here $\delta_{\text{eq}}$ is
an initial density fluctuation value at the end of radiation
dominated epoch.  In \cite{BBM97} the core is produced for
spherically symmetric clump by inverse flow caused by
annihilation of DM particles.  We shall show here that these
phenomena are not the main effects and that much stronger
disturbance of the velocity field in the core is produced by
tidal forces.  These forces originate due to non-sphericity of
the considered perturbation and presence of other fluctuations
nearby, including a fluctuation of larger scale in which the
considered fluctuation can be submerged.

The core formation in a fluctuation begins at the linear stage of
evolution and continues at the beginning of non-linear stage. The
tidal forces diminishes with time $t$ (see Eq.~(\ref{tlin})
below), while duration of this phase is proportional to $t$. Once
the core is produced it is not destroyed in the evolution
followed.  The stage of the core formation continues
approximately from $t_{\text eq}$ to the time of maximal
expansion $t_s$ and a little above, when a clump is detached from
expansion of universe and evolves in the non-linear regime.  Soon
after this period, a clump enters the hierarchical stage of
evolution, when the tidal forces can destroy it, but surviving
clumps retain their cores.

Let us expand the gravitational potential in the series near the
maximum of the density fluctuation taken as $\vec r=0$ at
arbitrary time $t$ during the {\em linear} growth of density
perturbations:
\begin{equation}
\phi(\vec r,t)= \phi_0\!+\!\left.\frac{\partial \phi}{\partial
r^i}\right|_0\!\!r^i
\!+\!\frac{1}{6}\left.\Phi_{ll}\right|_0\delta_{ij}r^ir^j
\!+\!\frac{1}{2}\left.T_{ij}\right|_0r^ir^j\!+\!\ldots\!,
\label{series}
\end{equation}
where
\begin{equation}
\Phi_{ij}=\frac{\partial^2 \phi(\vec r)}{\partial r^i\partial
r^j}, \quad T_{ij}=\Phi_{ij}-\frac{1}{3}\Phi_{ll}\delta_{ij}.
\end{equation}
The first term of series in Eq.~(\ref{series}) does not influence
the particle motion. The second term is zero as a condition of
maximum density. The third term describes the spherically
symmetric part of the potential (including the potential of the
homogeneous background) and also the perturbation potential. It
governs the radial motion of the particles. According to the
Poisson equation one has
\begin{equation}
\Phi_{ll}=\Delta\phi(\vec r)=4\pi G\bar\rho(1+\delta(\vec r)).
\label{puass}
\end{equation}
Finally, the fourth term, which contains the traceless matrix
$T_{ij}$, describes the tidal forces. They disturb the radial
motion of the particles and result in production of the core.

We shall start with definitions and notation. We assume that
density perturbations $\delta(\vec r)$ are Gaussian with a power
spectrum $P(k)$:
\begin{equation}
\langle\delta^*_{\vec k}\delta_{\vec k'}
\rangle=(2\pi)^3P(k)\delta_{\text{D}}^{(3)}(\vec k-\vec k'), \quad
\delta_{\vec k}=\int\delta(\vec r)e^{i\vec k\vec r}\,d^3r,
\label{pdef}
\end{equation}
where $\delta_{\text{D}}^{(3)}(\vec k-\vec k')$ is the Dirac
delta-function and angle brackets corresponds to ensemble
averaging. The power spectrum $P_{\text{eq}}(k)$ at the time
$t_{\text{eq}}$ is connected with the primordial power spectrum
$P_p(k)$ (at the epochs before the  horizon crossing) by relation
$P_{\text{eq}}(k)=P_p(k)T^2(k)$, where $T(k)$ is the transfer
function for cold dark matter (see e.~g. \cite{bardeen}).

From Eq.~(\ref{puass}) it follows that the power spectrum
$P_{\Phi}(k)$ of potential perturbations is related to $P(k)$ as
\begin{equation}
 P_{\Phi}(k)=(4\pi)^2G^2{\bar\rho}^2k^{-4}P(k).
 \label{ppph}
\end{equation}
Let us introduce the moments of spectrum $P(k)$
\begin{equation}
 \sigma_{(j)}^2=\frac{1}{2\pi^2}\int\limits_0^\infty k^2\,dk\,P(k)k^{2j},
 \label{sigj}
\end{equation}
and the similar moments $s_{(j)}^2$ for the perturbation field of
the gravitational potential. Calculating the divergent moments
for given mass $M$ we assume smoothing procedure of
\cite{bardeen}.

From Eq.~(\ref{ppph}) it follows that
\begin{equation}
s_{(j)}^2=(4\pi)^2G^2{\bar\rho}^2\sigma_{(j-2)}^2
\end{equation}
for $j\ge2$. Let us define $\zeta_{ij}=\partial^2 \delta(\vec
r)/\partial r^i\partial r^j$. Then according to \cite{bardeen},
its mean value over the ensemble is
\begin{equation}
\langle\zeta_{ij}\zeta_{kl}\rangle=\frac{\sigma_{(2)}^2}{15}
(\delta_{ij}\delta_{kl}+\delta_{ik}\delta_{jl}+\delta_{il}\delta_{jk}),
\end{equation}
which results in
\begin{equation}
\langle T_{ij}T_{ji}\rangle=\frac{2}{3}s_{(2)}^2=
\frac{2}{3}(4\pi)^2G^2{\bar\rho}^2\sigma_{(0)}^2, \label{ttsr}
\end{equation}
(in the following we shall use a notation $\sigma \equiv
\sigma_{(0)}$). Let us introduce the important physical quantity
$\nu$, the height of the peak density in units of dispersion (the
peak-height):
\begin{equation}
\nu = \delta_{\text{eq}}/\sigma_{\text{eq}}(M), \label{nu}
\end{equation}
where $\sigma_{\rm eq}(M) \equiv \sigma(t_{\rm eq},M)$.

After introduction of these quantities we shall come over to
calculation of DM particle velocities and core radius. The
velocity $\vec v(t)$ is given by sum of radial velocity $\vec
v_{\text{rad}}$ and an additional velocity $\vec v_{\text{tid}}$,
which will be considered as a small correction caused by tidal
forces. The radial velocity will be calculated without tidal
interaction taking into account from the equation
\begin{equation}
\frac{d\vec{v}_{\rm rad}}{dt}=-{\rm\bf grad}\, \phi (r),
\end{equation}
where spherically symmetric potential $\phi(r)$ is given by the
third term in r.h.s of Eq.~({\ref{series}). This equation
determines the radial motion of the particle, and its solution is
given in the parametric form as \cite{svvbm}:
\begin{equation}
r=r_s\cos^2\theta, \quad
\theta+\frac{1}{2}\sin2\theta=\frac{2}{3}\left(\frac{5\delta_{\text{eq}}}{3}
\right)^{3/2}\frac{t-t_s}{t_{\text{eq}}}. \label{param}
\end{equation}
The moment of maximum clump expansion $t_s$ and the distance
$r=r_s$ at this moment are
\begin{equation}
\frac{t_s}{t_{\text{eq}}}= \frac{3\pi}{4}
\left(\frac{5\delta_{\text{eq}}}{3}\right)^{-3/2}, \quad
\frac{r_s}{r_{\text{eq}}}=\frac{3}{5\delta_{\text{eq}}},
\label{tsti}
\end{equation}
where $\delta_{\text{eq}}$ is the initial fluctuation value (at
$t=t_{\rm eq}$).

Tidal forces give rise to the additional velocity
$\vec{v}_{\text{tid}}$. Its evolution is described by equation
\begin{equation}
\frac{dv_{\text{tid,i}}}{dt}=-T_{ij}(t)r^j \label{vshev}
\end{equation}
where in the linear approximation
\begin{equation}
T_{ij}(t)=T_{ij}(t_{\text{eq}})\left(\frac{t}{t_{\text{eq}}}\right)^{-4/3}.
\label{tlin}
\end{equation}
The linear approximation for tidal forces is justified because
they are generated by the large-scale perturbations which become
nonlinear later than the small-scale perturbation under
consideration.

Now we find the value $\vec v_{\text{tid}}$ at the moment when
the density nonlinearity sets in, $\delta\simeq1$, or more
exactly at the moment of a maximum expansion $t_s$. After
integration of (\ref{vshev}) with the help of Eq.~(\ref{param})
we obtain
\begin{equation}
v_{\text{tid,i}}(t_s)=
18^{1/3}\left(\frac{5\delta_{\text{eq}}}{3}\right)^{1/2}t_{\text{eq}}
f(\delta_{\text{eq}})T_{ij}(t_{\text{eq}})r^j(t_s), \label{vtid}
\end{equation}
where the function
\begin{equation}
f(\delta_{\text{eq}})=\frac{2}{3}\int\limits_
{(5\delta_{\text{eq}}/3)^{1/2}}^{\pi/2}
d\phi\left(\phi-\frac{1}{2}\sin2\phi\right)^{-4/3}\sin^4\phi
\label{f1}
\end{equation}
is plotted in the Fig.~\ref{fig1}.
\begin{figure}
\includegraphics[width=8.5cm]{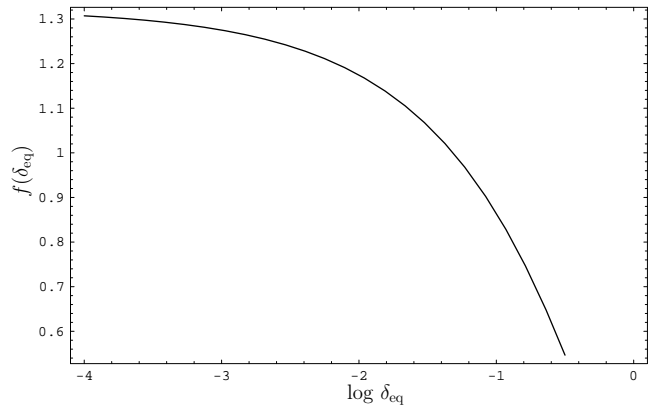}
\caption{Function $f(\delta_{\text{eq}})$ defined by
Eq.~(\ref{f1}). The ratio of the core radius to the clump radius
in typical case $\simeq0.3\nu^{-2}f(\delta_{\text{eq}})$.}
\label{fig1}
\end{figure}
We may use approximately $f(\delta_{\text{eq}})\simeq1$ for the
values of $\delta_{\text{eq}}$ at interest (asymptotically
$f(\delta_{\text{eq}}) \to 1.32$ at $\delta_{\text{eq}}\to0$).

To find the core radius $R_c$ of the clump we shall use the method
similar to that in \cite{ufn1}. Since $\mathbf{rot}\,\vec
v_{\text{tid}}=0$ and $\mathbf{div}\,\vec v_{\text{tid}}=0$, the
tensor $T_{ij}$ has the following diagonal form in the coordinate
system connected with the main axes.

\begin{equation}
T'_{ij}=\left(
\begin{array}{ccc}
{\displaystyle \tau } & & \\ & {\displaystyle \tau } & \\ & &
{\displaystyle -2\tau }
\end{array}
\right). \label{eps3}
\end{equation}
The value $\tau$ from Eq.~(\ref{eps3}) is connected with the core
radius $R_c$ due to energy relation $\Delta E\simeq\Delta V$,
where $\Delta E\simeq\int d^3r
\rho_{\text{int}}(t_s)v_{\text{tid}}^2(t_s)/2$ is the work
performed by tidal forces, and $\Delta V\simeq GMM_c/R$, where
$M_c\sim4\pi R_c^3\rho_c/3$, is potential energy of the core. It
gives for the relative core radius
\begin{equation}
x_c\simeq
\frac{2^{2/3}3^{10/3}}{\pi}\frac{\tau^2t_{\text{eq}}^2\delta_{\text{eq}}}
{G\bar\rho_{\text{int}}} f^2(\delta_{\text{eq}}). \label{core1}
\end{equation}
From invariance of the matrix trace relative to the change of
coordinates one has
\begin{equation}
\tau^2=T'_{ij}T'_{ji}/6=T_{ij}T_{ji}/6. \label{vpsttt}
\end{equation}
Since the correlator
\begin{equation}
\langle T_{ij}\delta\rangle=
\langle(\Phi_{ij}-\frac{1}{3}\Phi_{ll}\delta_{ij}) (\Phi_{ll}(4\pi
G\bar\rho)^{-1}-1)\rangle=0,
\end{equation}
the quantities $\delta$ and $T_{ij}$ are statistically
independent and we may average them independently. Averaging
$\langle\tau^2\rangle$ over the tidal force field $T_{ij}$ with
help of Eqs.~(\ref{ttsr}) and (\ref{vpsttt}), we obtain the main
result of this section for the relative core radius:
\begin{eqnarray}
 x_c=\frac{R_c}{R}&\simeq&
 \frac{\pi2^{5/3}3^{13/3}}{5^3}
 G\rho_{\text{eq}}t_{\text{eq}}^2\nu^{-2}f^2(\delta_{\text{eq}}) \nonumber \\
  &\simeq&
 0.3\nu^{-2}f^2(\delta_{\text{eq}}),
 \label{coreit}
\end{eqnarray}
where $\nu$ is given by Eq.~(\ref{nu}).

The fluctuations with  $\nu \sim 0.5 - 0.6$ have $x_c \sim 1$,
i.e. these fluctuations are practically destroyed by tidal
interactions. Most of galactic clumps are formed from $\nu \sim
1$ peaks. As it will be demonstrated in Sec.~\ref{res}, those
clumps that survive till now are characterized by $\nu\simeq1.6$,
but the main contribution to the annihilation signal is given by
the clumps with $\nu\simeq2.5$ for which $x_c\simeq0.05$.

In an alternative approach for calculation of the core radius one
may define $R_c$ as the minimum deviation of a typical particle
trajectory from a center of the clump.  After clump virialization
a particle at the average distance $R/2$ from the center has an
angular momentum $m v_{\rm tid}R/2$.  At $t>t_s$ the tidal force
is already small and angular momentum is approximately
concerved.  Then at the minimum distance from the core, $R_c$,
one has $R_c V\sim v_{\rm tid}R/2$, where $V=(2GM/R)^{1/2}$ is
the velocity of typical particle in the center of the virialized
clump.  Calculating $v_{\rm tid}$ from Eq.~(\ref{vtid}) and using
Eq.~(\ref{ttsr}) one obtains
\begin{equation}
 x_c\simeq
 0.15\nu^{-1}f(\delta_{\text{eq}}),
 \label{coreit1}
\end{equation}
which numerically is very close to Eq.~(\ref{coreit}) for typical
values of $\nu\simeq1-3$.}

The core radius, given by the Eq.~(\ref{coreit}), is much larger
than $x_c \simeq \delta_{\rm eq}^3$ obtained in \cite{ufn1} and
$x_c$ from \cite{BBM97}. The core radius found in \cite{BBM97} is
valid only in a perfectly symmetric case.

Several remarks are in order.

Tidal forces prevent the appearance of singularity during
evolution of the clump, but if such singularity has somehow
appeared, tidal interactions cannot destroy it.

In the calculations above we operated with average tidal force,
described by Eq.~(\ref{ttsr}). In reality this force fluctuates
due to positions of surrounding fluctuations which can overlap
with the considered one or to be far away from it. As a result
some clumps can be destroyed and those survived have different
core radii $R_c$. This effect increases the annihilation signal.

We assume that DM distribution within the core is much flatter
than $r^{-\beta}$. Between $R_c$ and the beginning of the
asymptotic regime $\rho_{\rm int} \propto r^{-\beta}$ there is a
transition zone. During the hierarchical evolution (see next
section) this zone expands due to tidal interaction in the
hierarchical structures. However, this interaction cannot destroy
the existing core.

The above calculations are also valid for the formation of
galaxies. It is useful to compare Eq.~(\ref{coreit}) with
observations of galaxies. In \cite{bardeen} the number of peaks
in the Gaussian random field is confronted with the observed
density of spiral and elliptical galaxies. It was found that
these galaxies are formed mostly from peaks with $\nu\simeq3$.
According to Eq.~(\ref{coreit}) for these peaks $x_c\simeq0.033$.
The rotation curves in the central part of dwarf and Low Surface
Brightness galaxies  are measured \cite{burkert,blok} and
constant-density cores were revealed. Existing observations do not
contradict the presence of extended uniform core with radius
$R_c=x_cR\sim3.3$~kpc in spiral and elliptical galaxies. However,
at these distances the baryonic matter dominates, which makes the
observation of DM core even more difficult. The extreme value
$R_c=R$ corresponds to $\nu_{\text{min}}\approx 0.55f$. The
Gaussian peaks with these $\nu$ are completely washed out by
tidal forces and do not produce the gravitationally bound
objects. The intermediate case $\nu\simeq1$ corresponds to most
numerous dwarf and irregular galaxies with the pronounced
overdensity in the central core.

\section{Tidal Destruction of Clumps in the Hierarchical Model}
\label{razrush}

In this section we shall study the destruction of clumps by the
tidal interaction which occurs at the formation of the
hierarchical structures, but long time before the galaxy
formation. This interaction arises when two clumps pass near each
other and when a clump moves in the external gravitational field
of the bigger object (host) to which this clump belongs.  In both
cases a clump is exited by the external gravitational field,
i.~e.  its constituent particles obtain additional velocities in
the c.~m.  system.  The clump is destroyed if its internal energy
increase $\Delta E$ exceeds the corresponding total energy $|E|
\sim GM^2/2R$.  In Sec.~\ref{excitation} we shall calculate the
rate of excitation energy production due to both aforementioned
processes. Respectively in Sec.~\ref{hierarchy} we shall
calculate the survival probability for a clump in the
hierarchical model, in which the smaller clumps are embedded in
the bigger one, and the latter enters into a more bigger clump
{\sl etc}.  But first we shall describe the necessary
generalities and definitions.

\subsection{Generalities and definitions}
\label{general}

The formation of DM objects with a fixed mass $M$ at the linear
stage is distributed over formation epochs $t_f$. In the
spherical model of the Press-Schechter theory \cite{ps74,cole}
the density perturbation at the epoch of object formation is
equal to $\delta_c=3(12\pi)^{2/3}/20\simeq1.686$:
\begin{equation}
\delta(M,t_f)=\delta_c \label{fcrit}
\end{equation}
Eq.~(\ref{fcrit}) gives the {\em formation criterion} for DM
objects. The formation criterion alone does not determine $t_f$
for a given mass $M$, because $\delta(M)$ has the Gaussian
distribution. The formation time $t_f$ can be fixed by an
additional condition, e.~g. $\nu=1$ (see Eq.~(\ref{nu})). The DM
objects which satisfy the formation criterion (\ref{fcrit}) and
$\nu=1$, or equivalently $\sigma(M,t_f)=\delta_c$, will be
referred to as the {\em typical} objects. For a given mass $M$
they are characterized by a fixed epoch of formation $t_f$. In
some parts of our consideration we shall simplify the problem,
assuming that clumps are {\em typical} instead of taking into
account their distribution over $t_f$.

We confine ourselves here only to the power-law spectrum
 of fluctuations
$P_{\text{eq}}(k)\propto k^n$, in which case
\begin{equation}
\sigma_{\text{eq}}(M)\propto M^{-(n+3)/6} \label{spn36}.
\end{equation}
The effective power-law index $n$ obtained from the
 expression above can be given as
\begin{equation}
n=-3-6\frac{\partial \ln \sigma_{\text{eq}}(M)}{\partial \ln M}.
\label{neff}
\end{equation}
In the case of an arbitrary $P(k)$ spectrum, the
$\sigma_{\text{eq}}(M)$ has also an arbitrary dependence on $M$.
Eq.~(\ref{neff}) defines $n_{\rm eff}$ for a some mass interval
$\Delta M$. In realistic cases (see e.~g. Eq.~(\ref{A5})),
Eq.~(\ref{neff}) is a good approximation because a power-law
index $n$ depends only weakly on $M$.

We shall introduce for convenience the characteristic values of
typical objects labeled by index `$\Lambda$' and described by the
condition $\sigma(t_\Lambda,M_\Lambda)=\delta_c$ at some fixed
moment $t_\Lambda$ or redshift $z_\Lambda$. We can choose these
quantities arbitrarily (because they will not enter into the
final results), but satisfying the condition $z_\Lambda\gg
(\Omega_{\Lambda}/\Omega_{m})-1$, e.~g. taking
$z_\Lambda\sim5-10$. The convenience of these normalizing values
is that at $t<t_{\Lambda}$ the $\Lambda$-term can be neglected,
while small-scale object formation occurs at these epochs. Let us
introduce also the dimensionless mass $m$ as
$$
  m=M/M_{\Lambda}.
$$
Using the formation criterion (\ref{fcrit}) we obtain
\begin{equation}
\delta(M,t_f)=\delta_{\text{eq}}\frac{1+z_{\text{eq}}}
{1+z_f}=\delta_c
\end{equation}
because the growth-factor for the rising mode
$D_g(t)\propto(1+z)\propto t^{2/3}$ in the standard cosmological
model at the matter dominated epoch. For a single clump with mass
$M$ obeying $\delta_{\text{eq}} = \nu\sigma_{\text{eq}}(M)$ with
an arbitrary $\nu$ one has
\begin{equation}
\bar\rho_{\text{int}}=\kappa\bar\rho(z_f)=
\kappa\rho_{\text{eq}}\left(\frac{1+z_{\text{eq}}}{1+z_{f}}\right)^3=
\kappa\rho_{\text{eq}}\frac{\nu^3\sigma_{\text{eq}}^3(M)}{\delta_c^3},
\label{rhocl}
\end{equation}
and
\begin{equation}
R=\left(\frac{3M}{4\pi\bar\rho_{\text{int}}}\right)^{1/3}=
\nu^{-1}R_\Lambda m^{(n+5)/6)}, \label{rad1}
\end{equation}
where $M$ and $R$ are respectively mass and radius of the clump,
$\kappa = 18\pi^2 \simeq 178$ and $R_{\Lambda} =
(3M/4\pi\kappa\bar\rho(t_{\Lambda}))^{1/3}$. The value of $\kappa$
describes the clump density increasing during contraction and it
can be found from Eq.~(\ref{param}) (see also \cite{cole}).

A number density of unconfined (free) clumps (i.e. of those not
belonging to the more massive objects) is given by the
Press-Schechter formulae \cite{cole}
\begin{eqnarray}
  \phi_{\text{PS}}(t,M)\,dM  &&   \nonumber \\
 =\!\left(\frac{2}{\pi}\right)^{1/2}\!\!\!\frac{\rho}{M}
 \frac{\delta_c}{D_g(t)\sigma_{\text{eq}}^2}
 \frac{d\sigma_{\text{eq}}}{dM} \times
 \exp\left[\frac{-\delta_c^2}{2D_g(t)^2\sigma_{\text{eq}}^2}\right]dM,
  && \nonumber \\
 \label{ps1}
\end{eqnarray}
where the growth-factor $D_g(t)$ is normalized as
$D_g(t_{\text{eq}})=1$. Let us consider the Press-Schechter
distribution for the clumps hosted by the larger clump at the
moment $t_f$ of its formation. Taking into account the density
increase factor $\kappa$ we obtain
\begin{equation}
 \psi_{\text{PS}}(t_f,m)dm=\kappa\phi_{\text{PS}}(t_f,m)dm
 \label{ps2}.
\end{equation}
As it will be demonstrated in this section the clumps are
destroyed by tidal interaction and each clump has a small
surviving probability $\xi < 0.01$. A survived clump during its
lifetime is surrounded by other clumps with distribution given by
Eq.~(\ref{ps2}). When a host clump is destroyed, the survived
clump finds itself hosted by larger clump with the small clump
distribution inside given by the same Eq.~(\ref{ps2}) but with
larger $t_f$. Since the tidal destruction is most effective at
small distances, one should always consider the smallest possible
host clump from the hierarchical structure, and the distribution
of the small clumps around one at the consideration is always
given by Eq.~(\ref{ps2}). The characteristic time is the time of
formation $t_f$ of the smallest host clump, though time of
destruction can be much larger than $t_f$.

The total energy (kinetic and potential) of a clump is given by
\begin{equation}
|E|=\frac{3-\beta}{2(5-2\beta)}\frac{GM^2}{R}. \label{Etot}
\end{equation}

\subsection{The rate of internal energy growth}
\label{excitation}

Consider a host clump with mass $M_h$, radius $R_h$, and with the
small clumps inside having distribution (\ref{ps2}) and moving in
the common gravitational potential with a velocity dispersion
$\sim V_h\simeq GM_h/R_h$. Interacting tidally with its
neighbors, a small clump increases its internal energy. We
calculate first the rate of internal energy increase due to these
interactions. The mass of the considered clump is $M=mM_\Lambda$,
it is characterized by an arbitrary $\nu$ and its interaction
with a target clump occurs at the impact parameter $l$. A target
clump is characterized by mass $M'$, radius $R'$, radius of the
core $R'_c=x_cR$ with $x_c \sim 0.1$ and by universal density
distribution (\ref{rho}).

An increase of internal energy of a clump $M$ during one fly-by
in the momentum approximation \cite{gnedin1} is given by:
\begin{equation}
\Delta E=\frac{1}{2}\int d^3r\,\rho(r)(v_x-\tilde v_x)^2,
\label{tidde}
\end{equation}
where $v_x$ is the velocity {\em increase} of DM particle in
direction of axis $x$, and $\tilde v_x$ is the same for c.~m. of
the clump. The axis $x$ connects the c.m.'s  of two clumps when
the distance between them is the shortest. One approximately has
\begin{equation}
v_x-\tilde v_x\simeq\frac{\partial v_x}{\partial l}\Delta l=
\frac{\partial v_x}{\partial l}r\cos\psi, \label{tiddvx}
\end{equation}
where $\psi$ is the polar angle in spherical coordinates.

For nearly straightforward propagation, the angle between $\vec
v_{\text{\text{rel}}}$ and the line connecting c.m.'s of the
clumps evolves as
\begin{equation}
\frac{d\phi}{dt}=-\frac{v_{\text{rel}}}{l}\cos^2\phi,
\end{equation}
where $v_{\text{rel}} \simeq\sqrt{2}V_h$. Changing variable $t$
to $\phi$ in the Newton equation one gets
\begin{equation}
\frac{dv_x}{d\phi}=-\frac{GM'(r'(\phi))}{v_{\text{rel}}l}\cos\phi,
\end{equation}
and after integration of this equation:
\begin{equation}
v_x=\frac{2GM'}{v_{\text{rel}}R'}g(y),
\end{equation}
where $y=l/R'$,
\begin{equation}
g(y)\!=\!\!\left\{
\begin{array}{ll}
\! y^{-1}, \qquad\qquad\qquad  y>1;  \\
\!
\left[1\!+\!y^{3-\beta}(1\!-\!y^2)^{1/2}~
\!\!_2F_1\left(\frac{3-\beta}{2},
\frac{1}{2},\frac{3}{2},1\!-\!y^2\right)\right.  \\
\! -(1\!-\!y^2)^{1/2}\Big]\big/y,   \quad y<1, &
\end{array}
\right.
\end{equation}
and $~_2F_1$ is the hypergeometric function.

The rate of internal energy growth due to collisions with all
other clumps is
\begin{equation}
\dot E=\int 2\pi l v_{\text{rel}}dl \int dM'\psi(M',t)\Delta E.
\label{dee1}
\end{equation}
After simple calculations, one obtains
\begin{eqnarray} \label{dee2}
 \dot E&=&\frac{4\pi(3-\beta)}{3(5-\beta)}
 \frac{G^2MR^2}{v_{\text{rel}}}
 \int\limits_{M}^{M_h}dM'{M'}^2\psi(M',t)
 \nonumber
\\
&& \times \left[ \frac{\lambda}{{R'}^2}+
\frac{1}{2}\left(\frac{1}{{R'}^2}-\frac{1}{R_h^2}\right) \right],
\end{eqnarray}
where
\begin{equation}
\lambda=\int\limits_{x'_c}^1dy~y\left(\frac{dg}{dy}\right)^2=0.11
\end{equation}
for $x'_c=0.1$ and dependence of $\lambda$ on $x'_c$ is very weak.

As the second process of tidal destruction we shall consider the
interaction with the common gravitational potential of a host
clump. Energy gain per mass unit at a distance $r$ from c.~m. of
the considered small clump $m$ during one periastron passage
\cite{gnedin1} is
\begin{equation}
\langle E_p\rangle=\frac{GM_h}{R_h^3}r^2
\left(\frac{R_h}{R_p}\right)^{\beta}
\chi_{\text{\text{ecc}}}(e)A(\omega\tau),
\end{equation}
where $e$ is the eccentricity, the function $\chi_{\text{ecc}}$
presented in \cite{gnedin1}, the adiabatic correction
$A(x)=(1+x^2)^{-\gamma}$, $\gamma\simeq2.5 - 3$ and $R_p$ is the
periastron separation. The energy gain of a clump during one
period $T_{\text{\text{orb}}}$ is $\Delta E=\int\langle
E_p\rangle\rho_{\text{int}}(r)d^3r$, and the rate of energy
growth is
\begin{equation}
\dot E=\frac{2\Delta E}{T_{\text{\text{orb}}}}. \label{deorb}
\end{equation}
The rate of energy growth due to the both of aforementioned
processes is given by the sum of (\ref{dee2}) and (\ref{deorb}).
Using the distribution (\ref{ps2}) in the integral of
(\ref{dee2}) and the total energy of a clump given by
Eq.~(\ref{Etot}) we find:
\begin{equation}
\frac{1}{T(m,m_h,\nu,\nu_h)} \equiv \frac{\dot E}{E}\simeq
2t_\Lambda^{-1}\mu\nu_h^{9/2}\nu^{-3} m^{(n+3)/2}
m_h^{-3(n+3)/4}\label{1t1},
\end{equation}
where
\begin{eqnarray}
\mu&=&\frac{2^{1/2}\kappa^{1/2}(5-2\beta)}{3(5-\beta)} \left[
\left(\frac{R_h}{R_p}\right)^{\beta}A(\omega\tau)\chi_{\text{ecc}}(e)
\right.
\nonumber \\
&& \left.+\frac{1}{4\pi^{1/2}} \left|\frac{n+3}{n+1}\right|
\left(2\lambda+\frac{n+5}{n+9} \right) \right]. \label{f2n1}
\end{eqnarray}
The first term in the square brackets describes the interaction
with common gravitational field of the host clump, while the
second term --- ``collisions'' with small clumps inside the host
clump. Usually the former is larger than the latter. For
calculations we shall consider an average orbit with
$R_h/R_p\simeq2$ and $T_{\text{orb}}\simeq2R_h/V_h$, and put
$A(\omega\tau)\chi_{\text{ecc}}(e)\sim1$. If to neglect the tidal
interactions with the small clumps (the second term in the square
brackets in Eq.~(\ref{f2n1})) and to use $\beta=1.8$, one has
$\mu\simeq9.6$. The dependence of our final result (mass function
of the clumps) on $\mu$ is weak, approximately as $\mu^{-1/3}$.

\subsection{Survival probability in the hierarchical model}
\label{hierarchy}

A small clump with mass $m$ during its life time can be a
constituent part of many host clumps of successively larger
masses $m'$ and virial velocities $V'$ . After tidal disruption
of the lightest host, a small clump automatically becomes a
constituent part of the heavier host etc. Transition of a small
clump from one host to another occurs almost continuously in time
up to formation of a host where tidal destruction becomes
inefficient. A fraction of small clumps with mass $m$ escaping
the tidal destruction is given by $e^{-J}$, with
\begin{equation}
J\!\simeq\!\sum\limits_{m'}\! \frac{\Delta t}{T(m,m',\nu,\nu')},
\label{jsum}
\end{equation}
where summation goes over the intermediate big clumps which
successively host the small clump $m$, and $\Delta t$
approximately equals to the difference of formation times of two
successive hosts.

Let us introduce the notation: $m_1$ is the mass of the first
(lightest) host clump which contains the considered light clump
$m$, and $m_n$ is the last such object, e.~g. the Galactic halo.
A formation epoch for the host clump $m'$ is
\begin{equation}
 t_f(m',\nu') = t_\Lambda\left(\frac{1+z'}{1+z_{\Lambda}}\right)^{-3/2}
 =t_\Lambda m^{'(n+3)/4}
\nu^{'-3/2}.
 \label{sumint}
\end{equation}
Note that $J$ does not depend on $t_{\Lambda}$ since it enters
linearly in $t_f$ and in $T$ as seen from Eq.~(\ref{1t1}).

The first host gives a major contribution to the clump
destruction, especially if its mass $m_1$ is close to $m$ and
$\nu_1\simeq\nu$. For the considered clump $m$ and for the first
two hosts $\sigma_{\text{eq}}(M) \simeq
\sigma_{\text{eq}}(M_1)\simeq\sigma_{\text{eq}}(M_2)$ because of
$\sigma_{\text{eq}}(M)$ depends weakly on $M$ according to
(\ref{neff}). The considered clump $m$ and the first two hosts
differ mainly by masses and by values of $\nu$
($\nu\ge\nu_1\ge\nu_2$). It justifies the following
simplification: we consider all hosts beginning from the second
one as the typical objects ($\nu_i=1$ for $i=2,3,\ldots$) and
separating the first term in the sum (\ref{jsum}), substitute the
remaining sum by the integral.
\begin{eqnarray}
 J & \simeq & \frac{t_f(m_2,\nu_2=1)-t_f(m_1,\nu_1)}{T(m,m_1,\nu,\nu_1)}
 \nonumber
 \\
 &+&\int\limits_{m_2}^{m_n}\!\frac{1}{T(m,m',\nu,\nu'=1)}
 \frac{dt_f(m',\nu'=1)}{dm'}dm'.
 \label{sumint1}
\end{eqnarray}
We may change now the lower limit $m_2$ to $m_1$ in the above
integral because it depends on $m_2$ weakly, only through
$m_2^{(n+3)/2}$ with $(n+3)/2\ll1$, where $n$ is given by
Eq.~(\ref{neff}). In addition we may put the upper limit
$m_n\to\infty$ without loss of accuracy. Inserting
Eq.~(\ref{1t1}) into (\ref{sumint1}) we obtain finally the
following approximate expression for $J$:
\begin{eqnarray}
 J&\simeq&
 2\mu\frac{\nu_1^{9/2}}{\nu^3}
 \left(\frac{m}{m_1}\right)^{(n+3)/2}(1-\nu_1^{-3/2})\theta(\nu_1-1)
 \nonumber
 \\
 &+&\mu\nu^{-3}
 \left(\frac{m}{m_1}\right)^{(n+3)/2},
 \label{sumint2}
\end{eqnarray}
where the step-function $\theta(x)=1$ for $x>0$ and $\theta(x)=0$
for $x\leq0$. In the case of $\nu_1=1$ the formation moment of
the first host almost coincides with the formation moment of the
second one, for which $\nu_2=1$ in the used approach.

\begin{figure}
\includegraphics[width=8.5cm]{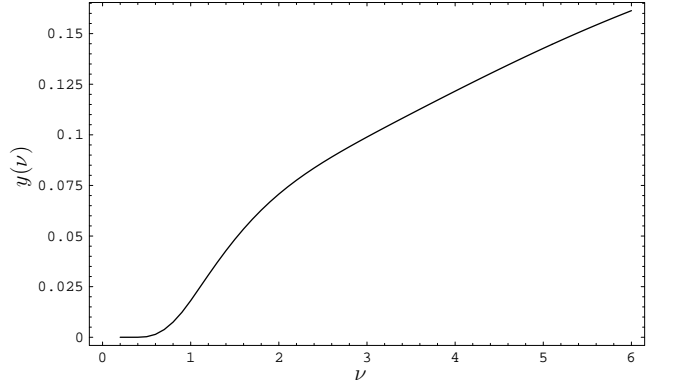}
\caption{Function $y(\nu)$ from Eq.~(\ref{psiitog}) obtained by
the numerical integration of (\ref{phiin}). This curve is valid
with good accuracy for all $\beta$ from the interval $1 \leq
\beta \leq 2$. }
 \label{figynu}
\end{figure}

The differential fraction of mass in the form of clumps which
escape the tidal destruction in the hierarchical objects can be
found as
\begin{eqnarray}
 \xi(n,\nu)\frac{dm}{m}d\nu&=&dm\,\,d\nu\,(2/\pi)e^{-\nu^2/2}
 \int\limits_0^{\nu}d\nu_1e^{-\nu_1^2/2}
 \nonumber \\
 \label{phiin}
 &&\!\times\!\int\limits_{t_f(m,\nu)}^{\infty}dt_1 \left|
 \frac{\partial^2 F(m,t_1)}{\partial m~\partial t_1}
 \right| e^{-J},
\end{eqnarray}
where the variable $m_1$ in $J$ is connected with $t_1$ and
$\nu_1$ according to Eq.~(\ref{sumint}), $F(m,t)$ is the mass
fraction of unconfined clumps with masses smaller than $m$ at the
moment $t$, which according to \cite{cole} is given by
\begin{equation}
 F(m,t)={\rm erf}\left\{
 \frac{\delta_c}{\sqrt{2}\sigma_{\text{eq}}(M)D_g(t)}\right\}.
\end{equation}
Here ${\rm erf(x)}$ is the error-integral and $D_g(t)$ is the
growth-factor. After numerical calculations for $\beta=1.8$ we
finally obtain:
\begin{equation}
 \xi(n,\nu)\simeq(2\pi)^{-1/2}e^{-\nu^2/2}
(n+3)y(\nu),
 \label{psiitog}
\end{equation}
where the function $y(\nu)$ is plotted in the Fig.~\ref{figynu}.
It weakly depends on $\beta$.

This is the main result of this section. We shall refer to $\xi$
as to the {\em clump survival probability}.

Our final result, the distribution of surviving clumps over their
masses $M$ and $\nu$, which for the fixed $M$ determines the
clump radius $R$, is given by $\xi(n,\nu)\, d\nu\, dM/M$, where
$\xi$ depends very weakly on $M$ only through the weak dependence
of $n(M)$ (see Eq. (\ref{neff})). For most numerous clumps with
$\nu=1$ and for the unit intervals $d\nu$ and $d\ln M$,~ $\xi$
has a meaning of fraction of DM mass in the form of clumps
relative to free DM particles, as it was introduced in
Eq.~(\ref{etasimp}). By integrating over  $\nu$, we obtain
\begin{equation}
\xi_{\text{int}}\simeq0.01(n+3). \label{xitot}
\end{equation}
This means that, for different $n$,  about $0.1-0.5$\% of clumps
survive the stage of tidal destruction in each logarithmic mass
interval $\Delta\ln M\sim1$.

Several remarks are in order.

The physical meaning of the survived clump distribution
$\xi(n,\nu)d\nu dM/M$ is different from that for unconfined
(free) clumps given by the Press-Schechter mass function
$\partial F/\partial m$. Surviving clump distribution implies
that each DM particle belongs simultaneously to several host
clumps put into each other, and by this reason the integral $\int
\xi dm/m$ is divergent.

To calculate $\xi$ we need to know the power-law index $n$ in the
perturbation spectrum, which can be taken as $n_{\rm eff}$ from
Eq.~(\ref{neff}) for given $\sigma_{\text{eq}}(M)$. To find the
latter, the primeval (e.~g. inflation) power spectrum of
fluctuation is needed. The simplest inflation models give
$P_p(k)\propto k^{n_p}$ with $n_p \approx 1$. The analysis of the
WMAP measurement of CMB anisotropy \cite{wmap} gives power law
spectrum with $n_p=0.99 \pm 0.04$ in good agreement with $n_p=1$.
However, when data from 2dF galaxy power spectrum and Ly-$\alpha$
are included in analysis, the best-fit favors a mild tilt,
$n_p=0.96 \pm 0.02$.

The variance $\sigma_{\text{eq}}(M)=\sigma_{(0)}$ from
Eq.~(\ref{sigj}) in the small scale range is found in
\cite{bugaev} (see also \cite{ssw}) . We present this result as
\begin{eqnarray}
 \sigma_{\text{eq}}(M)&\simeq&
 \frac{2\times10^{-4}}
 {\sqrt{f_s(\Omega_{\Lambda})}}
 \left[\ln\left(\frac{k}{k_{\text{eq}}}\right)\right]^{3/2}
 \left(\frac{k}{k_{h0}}\right)^{(n_p-1)/2} \nonumber
 \\
 &\simeq&8.2\!\times\!10^{3.7(n_p-1)-3}
 \!\left[1-0.06\lg\left(\frac{M}{M_{\odot}}\right)\right]^{3/2}
 \nonumber \\
 &\times&  \left(\frac{M}{M_{\odot}}\right)^{(1-n_p)/6},
 \label{A5}
\end{eqnarray}
where $k_{\text{eq}}$ and $k_{h0}$ correspond to mass inside the
cosmological horizon at the moments $t_{\text{eq}}$ and $t_{0}$,
respectively, and
\begin{equation}
f_s(\Omega_{\Lambda})=1.04-0.82\Omega_{\Lambda}+2\Omega_{\Lambda}^2
\end{equation}
according to \cite{bugaev}. We used above the relation $k\propto
M^{-1/3}$ and the values
\begin{equation}
M_{\text{eq}}\!=\!1.5\!\times\!10^{49}\Omega_m^{-2}h^{-4}\mbox{
g}, \ \ M_{h0}\!=\!6\!\times\!10^{55}\Omega_mh^{-1}\mbox{ g}.
\end{equation}

Using the power-law spectrum of fluctuations down to the small
scales, while normalization by the CMB anisotropy is performed at
large scale, implies an extrapolation of the spectrum by many
orders of magnitudes. Such extrapolation is justified only by the
confidence to inflation models which predict the power-law
spectrum valid for many orders of mass magnitudes.

It is interesting to note that the differential number density of
clumps in the Galactic halo $n(M)\,dM \propto dM/M^2$, obtained
from Eq.~(\ref{psiitog}), is very close (including the
normalization coefficient) to  that obtained in the numerical
simulations for clumps with large masses $M\geq10^8M_{\odot}$
[$n(M)dM \propto dM/M^{1.9}$ \cite{ghigna98}]. Strictly speaking
our calculations are not valid for clumps with these masses,
because of their destruction in the halo up to present epoch
$t_0$ and accretion of new clumps to the halo. Nevertheless, for
the small interval of masses where the power-law spectrum can be
used as a rather good approximation, our approach can be roughly
valid.

\section{Numerical Results}
\label{res}

In this section we shall present the numerical results of our
calculations: enhancement factor for annihilation signal, the
distribution of DM clumps over their masses $M$ and radii $R$,
and the distribution of clumps in the galactic halo.

Using Eqs.~(\ref{ihal}) and (\ref{hom}) we find the enhancement
of the annihilation signal due to clumpiness of the halo as
generalization of Eq.~(\ref{etasimp}) for the clumps distributed
over $M$ and $R$:
\begin{equation}
 \eta(M_{\text{min}},n_p)\!=\!
 1\!+\!\frac{1}{\tilde\rho_{\text{DM}}}\!\!\!
 \int\limits_{\nu_{\text{min}}}^{\infty}
 \!\!\!\!d\nu\!\!\!\!
 \int\limits_{M_{\text{min}}}\!\!\!\!\!
 \frac{dM}{M} S[\beta,x_c(\nu)]\xi
 \bar\rho_{\text{int}}(M,\nu),
 \label{itogexpr}
\end{equation}
where
\begin{equation}
\xi=\xi(n,\nu)
\end{equation}
is defined by Eq.~(\ref{psiitog}), an effective spectrum index
$n(M,n_p)$ is calculated from Eq.~(\ref{neff}) for primeval
(inflation) spectrum index $n_p$, with $\sigma_{\rm eff}(M)$
taken in the form (\ref{A5}); $\bar\rho_{\text{int}}$ is given by
Eq.~(\ref{rhocl}) and $\nu_{\text{min}}\simeq0.55$. Function $S$
is taken in the form (\ref{sbig}), which corresponds to the
density profile (\ref{rho}) and we used Eq.~(\ref{coreit}) for a
clump virial radius $R$.

The major part of the survived clumps are formed from
fluctuations with a mean value of the peak height
\begin{equation}
 \langle\nu\rangle\simeq
 \frac{\displaystyle \int d\nu\, \nu e^{-\nu^2/2}y(\nu)}
 {\displaystyle \int d\nu\, e^{-\nu^2/2}y(\nu)} \simeq1.6.
 \label{nueff0}
\end{equation}
Meanwhile the main contribution to the enhancement of the
annihilation signal (\ref{itogexpr}) comes from the clumps with
an effective value of the peak height:
\begin{equation}
 \langle\nu\rangle_{\text{ann}}\simeq
 \frac{\displaystyle
 \int\int d\nu\, \nu S\xi\bar\rho_{\text{int}}dM/M}{\displaystyle
 \int\int d\nu\, S\xi\bar\rho_{\text{int}}dM/M}\simeq2.5
 \label{nueff}
\end{equation}
for $\beta=1.8$. The clumps with $\nu\simeq2.5$ have
$x_c\simeq0.05$.

For the Galactic halo we use the NFW density profile \cite{nfw}:
\begin{equation}
 \rho_{\text{DM}}(l)=\frac{\rho_0}{ \displaystyle (l/L)
 \left(1+l/L)\right)^2}, \label{b1b2}
\end{equation}
with $L=45$~kpc according to \cite{berez2}, and $\rho_0$ fixed by
the local density value
$\rho_{\text{DM}}(r_{\odot})=0.3$~GeV~cm$^{-3}$. With these
parameters the halo mass within the virial radius of $100$~kpc is
$10^{12}M_{\odot}$. Eq.~(\ref{rhoeff}) gives
$\tilde\rho_H=1.02\rho_H(r_{\odot})$, i.~e. these values
practically coincide.

\begin{figure}
\includegraphics[width=8.5cm]{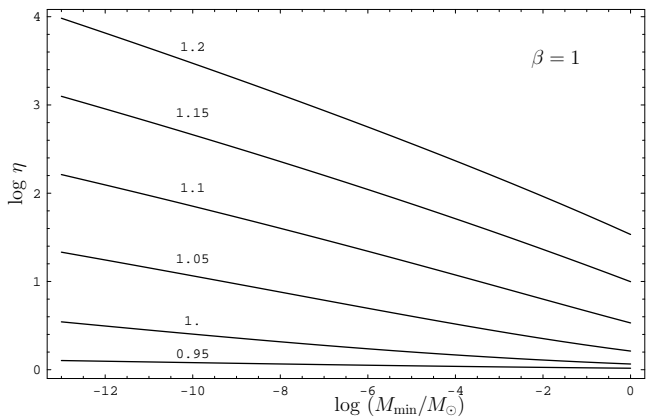}
\caption{The global enhancement $\eta$ of the annihilation signal
from Eq.~(\ref{itogexpr}) as a function of the minimal clump mass
$M_{\text{min}}$, for clump density profile with index $\beta=1$
and for different indices $n_p$ of primeval perturbation
spectrum. The curves are marked by the values of $n_p$.}
\label{bet1l}
\end{figure}

\begin{figure}
\includegraphics[width=8.5cm]{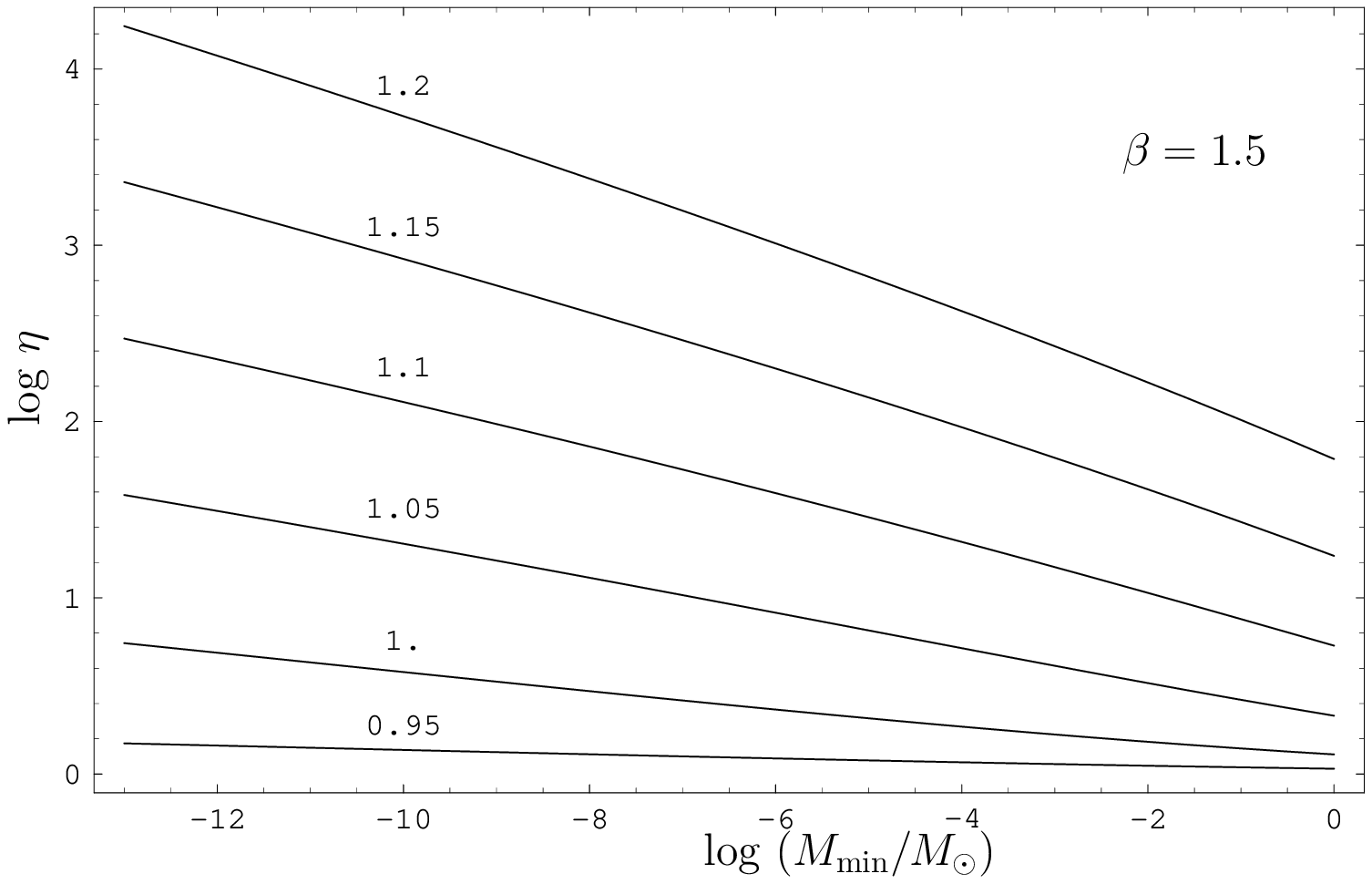}
\caption{The same as Fig.~\ref{bet1l} but for $\beta=1.5$.}
\label{bet15l}
\end{figure}

\begin{figure}
\includegraphics[width=8.5cm]{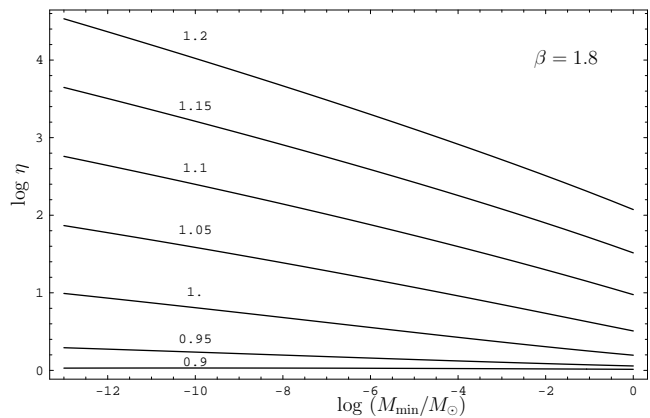}
\caption{The same as Fig.~\ref{bet1l} but for $\beta=1.8$.}
\label{bet18l}
\end{figure}

The values of global enhancement $\eta(M_{\text{min}},n_p)$ as
given by Eq.~(\ref{itogexpr}) are displayed in
Figs.~\ref{bet1l}--\ref{bet18l} for different values of
$M_{\text{min}}$, $\beta$ and $n_p$. As a representative example
consider the clump with the Gurevich-Zybin \cite{ufn1} density
profile with $\beta=1.8$ (see Fig.~\ref{bet18l}): numerically
$\eta=5$ for $M_{\text{min}}= 2\times 10^{-8}M_{\odot}$ and
$n_p=1.0$. It strongly increases at smaller $M_{\text{min}}$ and
larger $n_p$. For example, for $n_p= 1.1$ and $n_p=1.2$ at the
same $M_{\text{min}}=2\times 10^{-8}M_{\odot}$, enhancement
becomes tremendously large, $\eta=130$ and $\eta=4\times10^3$,
respectively.

Our approach is based on the hierarchical clustering model in
which smaller mass objects are formed earlier than the larger
ones, i.~e. $\sigma_{\text{eq}}(M)$ diminishes with the growing
of $M$. This condition is satisfied for objects with mass
$M>M_{min}\simeq2\times10^{-8}M_{\odot}$ only if the primordial
power spectrum has the value of the power index $n_p>0.84$. As
seen from Figs.~\ref{bet1l}--\ref{bet18l}, in this case the
enhancement of the annihilation signal in fact is absent,
$\eta\simeq1$, for $n_p<0.9$.

We discussed above the global enhancement of the annihilation
signal. In fact the enhancement varies in the different
directions relative to the Galactic Center (GC). It can be easily
seen from Eqs.~(\ref{ihal}) and (\ref{hom}), which show that
while $I_{\rm cl}$ is proportional to $n_{\rm cl}(l)$ (and thus
to $\rho_{\rm DM}(l)$ everywhere except the core), $I_{\rm hom}$
is proportional to $\rho_{\rm DM}^2(l)$. It implies that relative
contribution of $I_{\rm hom}$ increases in the directions close
to GC where $\rho_{\rm DM}(l)$ is larger. This effect is further
enhanced  by the destruction  of the clumps in the core around
the GC. Our numerical calculations confirm this expectation: the
ratio of enhancement in the directions to GC and Anticenter  is
0.2 for NFW density profile and for the core radius 3~kpc.
Therefore, the presence of the clumps diminishes the anisotropy
of the annihilation signal, caused by the density profile of DM
in the halo.

Our results suffer from uncertainties in input parameters. As was
mentioned above
$\tilde\rho_{\text{DM}}\simeq\rho_{\text{DM}}(r_{\odot})$. It
remains approximately true not only for the NFW density profile
but also for other profiles discussed in the literature. For
example, for isothermal profile with the core radius 10 kpc,
$\tilde\rho_{\text{DM}}=0.65\rho_{\text{DM}}(r_{\odot})$. Another
uncertainty in the value of $\eta$ is imposed by value of
$\rho_{\text{DM}}(r_{\odot})$. According to different estimates
$\rho_{\text{DM}}(r_{\odot})=(0.2 - 0.6)\mbox{~GeV~cm}^{-3}$. The
corresponding uncertainty in $\eta$ is given by factor $ 0.5 -
1.5$.

For illustration we shall describe numerically the properties of
clumps which give the main contribution to annihilation rate.
Basically, they are those with $M \sim M_{\rm min}$ and
$\nu\sim\langle\nu\rangle$. The r.m.s. fluctuation values
$\sigma_{\text{eq}}(M)$ for clumps with minimal mass
$M_{\text{min}}\simeq2\times10^{-8}M_{\odot}$ and for $n_p=1$ and
1.2 are $\sigma_{\text{eq}}=0.015$ and 0.14, respectively,
according to Eq.~(\ref{A5}). From Eq.~(\ref{nueff}) the effective
value of $\nu=\delta_{\text{eq}}/\sigma_{\text{eq}}$ is
$\langle\nu\rangle\simeq2.5$. From Eqs.~(\ref{rhocl}),
(\ref{rad1}) it follows that clumps with this $\nu$ are
characterized by the density and radius
$\bar\rho_{\text{int}}\simeq2\times10^{-22}$~g~cm$^{-3}$,
$R\simeq3.6\times10^{15}$~cm and $\bar\rho_{\text{int}}
\simeq2\times10^{-19}$~g~cm$^{-3}$, $R\simeq3.7\times10^{14}$~cm
for $n_p=1$ and 1.2, respectively. The part of Galactic halo mass
in the form of these clumps is of the order of
$\xi_{\text{int}}\sim 0.002$ according to Eq.~(\ref{xitot}). A
mean number density of the clumps in the halo is $\sim
25$~pc$^{-3}$.

We have given above the characteristic values for clumps with
dominant contribution to the annihilation signal. The general
distribution of clumps in the Galactic halo can be readily
calculated numerically from Eq.~(\ref{psiitog}), changing (for
given $M$) the distribution over $\nu$ by that over $dR$:
\begin{equation}
n_{\text{cl}}(M,R)d\ln M d\ln R =
\frac{\rho_{\text{DM}}(r_{\odot})}{M}\xi(M,\nu)\,d\ln M \, d\nu.
\label{n-MR}
\end{equation}
Note that definition of the clump number density $n_{\text{cl}}$
here does not coincides with the similar one in
Eq.~(\ref{nclfirst}) where $dN$ is given per $dR$ and $dM$.

\begin{figure}
\includegraphics[width=8.5cm]{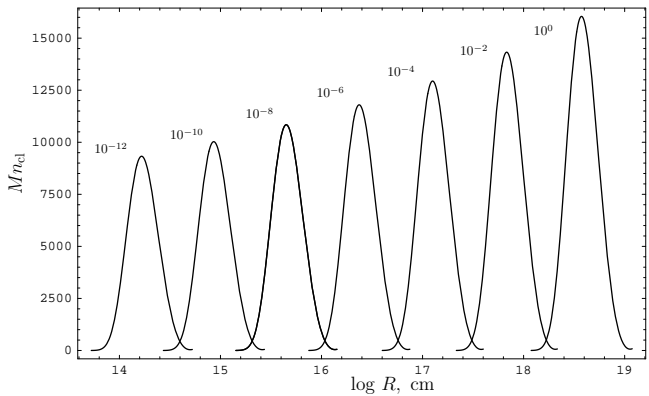}
\caption{The mass density of clumps  in the Galactic halo
$M\,n_{cl}(M,R)\,$ in units $M_{\odot}/$~kpc$^{3}$, from
Eq.~(\ref{n-MR}), as function of their radius $R$ at the distance
$8.5$~kpc from the Galactic center for $n_p=1.0$. The curves are
labeled by values of clump masses in $M_{\odot}$ } \label{figncl}
\end{figure}

The distribution $M\,n_{cl}(M,R)$ is presented in
Fig.~\ref{figncl} as function of $R$ for different $M$ and for
distance $8.5$~kpc from the Galactic center.

The radius of a clump $R$ in most general case is determined by
$M$ and $\nu$. Due to weak dependence of $\sigma_{\text{eq}}(M)$
on $M$, the radius of the clump $R(M)$ with sufficiently good
accuracy is proportional to $M^{1/3}$. From Eqs.~(\ref{rhocl})
and (\ref{rad1}) we have
\begin{equation}
R\simeq
1.5\times10^{16}\left(\frac{M}{10^{-6}M_{\odot}}\right)^{1/3}
\left(\frac{\nu}{2.5}\right)^{-1} \mbox{ cm}, \label{RM}
\end{equation}
for $n_p=1.0$. For $n_p=1.1$ the numerical factor in the
Eq.~(\ref{RM}) is $5.5\times10^{15}$~cm.

How clumps are distributed in the Galactic halo? One may expect
that this distribution is the same as distribution of the free DM
particles in the halo. This is true for large distances $l$ from
Galactic Center, while at small $l$ the tidal interaction with
stars results in the destruction of small clumps and in the
formation of the core with radius $L_c$ in the clump
distribution. The destruction of a clump propagating in the space
filled by stars has been studied in \cite{ufn1}. The destruction
is important only for clumps inside the bulge, i.e. at the
distance $l\le 3$~kpc from the Galactic Center. Clumps outside
the bulge at the distances $3\mbox{~kpc}\leq l\leq10\mbox{~kpc}$
can interact tidally with stars in the disk. But time of crossing
the disk is very small (in comparison with orbital period) and
this process is not important.

The number density of clumps outside the bulge is proportional to
the halo density, e.~g. to $\xi \rho_{\rm DM}(l)$ in the case of
the NFW distribution given by Eq.~(\ref{b1b2}), or can be
obtained from (\ref{b1b2}) and Fig.~\ref{figncl} by simple
scaling.

The density distribution for stars in the bulge
 according to \cite{lacost} is given by
\begin{equation}
\rho_s(l)=\left\{
\begin{array}{lr}
{\displaystyle \tilde\rho (l/\tilde r)^{-1.8}},  & l<\tilde r,
\\
\\
{\displaystyle \tilde\rho (l/\tilde r)^{-3}}, & l>\tilde r,
\end{array}
\right. \label{rhos}
\end{equation}
where $\tilde\rho=1.8M_{\odot}/$pc$^3$ and $\tilde r=800$~pc. From
Eq.~(46) of \cite{ufn2} by substituting Eq.~(\ref{rhos}) we
obtain that inside the bulge ($l\leq3$~kpc), the clumps with
$M\leq10^{-4}M_{\odot}$ are destructed during the Hubble time.
Thus, for these masses the core radius $L_c$ coincides with the
size of the bulge $L_{\rm bulge} \simeq 3$~ kpc.

\begin{figure}
\includegraphics[width=8.5cm]{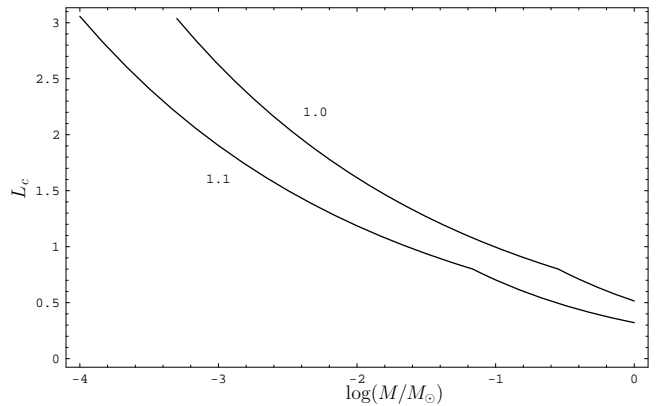}
\caption{The radius $L_c$ of the Galactic core (in kpc) in the
distribution of clumps with masses $M$ for $n_p=1.0$ and
$n_p=1.1$ in the case $\nu=2.5$.} \label{figtb}
\end{figure}

Clumps with $M\geq 10^{-4}M_{\odot}$ are destructed during the
Hubble time within distances from the Galactic Center shown in
Fig.~\ref{figtb} for $n_p=1.0$, $n_p=1.1$ and $\nu=2.5$. This
distance defines the radius of the core $L_c$ for clumps of the
given mass $M$.

Our calculations for enhancement of annihilation signal disagree
with those in \cite{BGZ,ABO,olinto}.

In \cite{BGZ} the singularity in the Galactic Center is cut at
very small core radius, which results in too strong annihilation
signal. According to our calculations the radius of the core is
much larger, and the distribution of the clumps in the halo also
has a core.

In \cite{ABO,olinto} the large enhancement of the signal is found
for heavy clumps with $M> 10^6 M_{\odot}$ \cite{ABO} and $M>10^2
M_{\odot}$ \cite{olinto}. If it were true, the total signal from
clumps with $M \geq M_{\rm min}$  would be too large. Too small
core radius was used in these calculations, too.

\section{Conclusions}

We have calculated the number density of the small-scale clumps
in the Galactic halo and their distribution over masses $M$,
radii $R$ and distances to the Galactic Center in the framework
of the standard cosmological model with the primeval density
perturbation $P(k) \propto k^{n_p}$ taken from the inflation
models with $n_p \simeq 1$ (the Harrison-Zeldovich spectrum). The
most important element of our calculations is inclusion of the
tidal interactions, which result in the formation of the clump
core and destruction of small-scale clumps.

We consider most conservative case of the Gaussian adiabatic
fluctuations which enter the non-linear stage of evolution,
$t_{\rm nl}$, at the matter-dominated epoch $t_{\rm nl}>t_{\rm
eq}$, where $t_{\rm eq}$ is the moment of equality. The time of
small-scale clump formation $t_f$ for a clump with mass $M$ is
given by two equation: formation criterion
$\delta(M,t_f)=\delta_c$ and by height of peak density of a
fluctuation in units of dispersion $\nu=\delta_{\rm
eq}/\sigma_{\rm eq}(M)$, taken at the epoch $t_{\rm eq}$ [see
Eqs.~(\ref{nu}) and (\ref{fcrit}) for explanation and notation].
All processes we are interested in, take place at $t \geq t_{\rm
eq}$ at the stage of non-linear evolution. The growth of
fluctuations in the non-linear regime we study in the framework
of the Press-Schechter theory of hierarchical clustering with the
tidal interactions included as the new element. The picture of
hierarchical clustering and clump destruction can be described in
the following way. The clumps of the minimal mass are formed
first. A clump of larger mass, which hosts the smaller clumps, is
formed later. A bigger clump, which includes the considered hosts
with their content, is formed further later etc. The clumps are
destroyed in tidal interactions with other small clumps and by
gravitational field of a host clump, with the former process
being subdominant. The calculated mass density of the survival
clumps, $\xi(M,\nu)d\nu dM/M$, with mass $M$ and $\nu$, is given
by Eq.~(\ref{psiitog}), with the survival probability being
typically $\xi \sim 0.001 - 0.005$. The clump number density in
the Galactic halo $n_{\rm cl}(M,R)$ for the clumps with mass $M$
and radius $R$ is shown in Fig.~\ref{figncl}. These clumps are
distributed in the Galactic halo as function of a distance to the
Galactic Center $l$. At large distance the distribution must be
the same as found in the numerical simulations (e.~g. the NFW
profile). At small distance there is a core produced by tidal
interaction of the clumps with the stars in the bulge. The radius
of the core, $L_c$ is given in Fig. \ref{figtb} at $M \geq
10^{-4}M_{\odot}$ and it is equal to the radius of the bulge $L_c
\sim 3$~kpc for smaller clump masses.

The mass spectrum  of the clumps is characterized by a cutoff at
$M_{\rm min}$. Its value depends on the properties of the DM
particle, and thus it is model dependent. The existing
calculations of $M_{\rm min}$ differ drastically: from $M_{\rm
min} \sim 10^{-12}M_{\odot}$ \cite{gzv12} to $M_{\rm min} \sim
10^{-7}M_{\odot}$ \cite{bino}.

Cold Dark Matter particles at high temperature $T>T_f \sim
0.05m_{\chi}$ are in thermodynamical (chemical) equilibrium with
cosmic plasma, when their number density is determined by
temperature. After decoupling at $t>t_f$ and $T<T_f$, the DM
particles remain for some time in the kinetic equilibrium with
plasma, when temperature of CDM particles $T_{\chi}$ is equal to
temperature of plasma $T$, but number density $n_{\chi}$ is not
Planckian any more. At this stage the CDM particles are not
perfectly coupled to the cosmic plasma. Collisions between a CDM
particle and fast particles of ambient plasma result in exchange
of momenta and a CDM particle diffuses in a space. Due to
diffusion, the DM particles leak from the small-scale
fluctuations, and thus their distribution obtain a cutoff at
minimal mass $M_D$. The diffusion coefficient is determined by
elastic scattering of DM particles off the plasma particles. Our
calculations made for neutralino, for which we have chosen the
pure bino state, give
\begin{equation}
M_{\text{D}}= 4.3\times\!10^{-13} \!\left(
\frac{m_{\chi}}{100\mbox { GeV}} \right)^{-15/8}
\!\!\left(\frac{\tilde M}{1\mbox{
TeV}}\right)^{-3/2}\!\!\!\!M_{\odot},
\end{equation}
where $m_{\chi}$ is the neutralino mass and $\tilde M$ is
(approximately) the mass of sneutrino and selectron, which are
assumed to be equal. The functional dependence of
Eq.~(\ref{dkdif}) and numerical value of Eq.~(\ref{mdif})
obtained in the diffusion approximation coincide with the
corresponding results obtained by different method in
\cite{gzv12}.

When the energy relaxation time for DM particles,
$\tau_{\text{rel}}$, becomes larger than the Hubble time
$H^{-1}(t)$, DM particles get out of kinetic equilibrium. This
conditions determines the time of kinetic decoupling $t_d$. At $t
\geq t_d$ CDM matter particles are moving in free streaming
regime and all fluctuations on the free-streaming scale
$\lambda_{fs}$ and less are washed out. In contrast to
\cite{bino}, we have calculated the free-streaming length
$\lambda_{fs}$ taking into account the distribution of neutralino
(bino) velocities over absolute values and angles from radial
directions. The cosmological expansion in the vicinity of $t_{\rm
eq}$ is taken exactly, without usual step-function approximation.
Our value of $M_{\rm min}$ due to free-streaming effect is
\begin{equation}
M_{\text{min}}=1.5\times10^{-8}\left(\frac{m_{\chi}}{100\mbox{
GeV}} \right)^{-15/8} \left(\frac{\tilde M}{1\mbox{
TeV}}\right)^{-3/2} M_{\odot};
\end{equation}
see Eq.~(\ref{mminnum}) for more parameters involved.

When normalized to the same masses of neutralino and sleptons,
our value of $M_{\rm min}$ coincides only by order of magnitude
with \cite{bino}.

The evolution of a density fluctuation in the non-linear regime
results in the density profile of a clump. The analytic theory of
this phenomenon was developed by Gurevich and Zybin (for a review
see \cite{ufn1}), for the numerical simulations see
\cite{nfw,moore1}. The initial single-stream flow leads to
formation of initial singularity. In contrast to
energy-dissipating matter (e.~g., baryons), in the flow of
non-dissipative matter the multistream instability develops
\cite{ufn1}, when at one point several streams with different
radial velocities exist. The surfaces with different number of
streams are separated by caustics, which number increases rapidly
towards the center. The matter is gravitationally captured in a
such structure. The density singularity is produced in the
center, unless the additional phenomena are included in the
consideration. As such the interaction with the damped mode
\cite{ufn1} and annihilation of DM particles \cite{BBM97} were
previously studied. We have demonstrated here that tidal forces
due to external gravitational field cause the deflection of DM
particles from radial motion, and prevent thus formation of
singularity. The produced core has a radius $R_c$ given in the
approximate form as
\begin{equation}
x_c=\frac{R_c}{R}\simeq 0.3\nu^{-2},
\end{equation}
(see Eq.~(\ref{coreit}) for the exact expression and the
discussion afterwards).  This radius is much bigger than those
obtained in \cite{ufn1,BBM97}.

The majority of clumps are formed from $\nu \sim 1$ peaks, while
the survived clumps are characterized on average by
$\nu\simeq1.6$.  The clumps which give the dominant contribution
to the annihilation signal have $\nu\simeq 2.5$.

In spite of small surviving probability, $\xi \sim 0.1 - 0.5 \%$,
clumps in most cases provide the dominant contribution to the
annihilation rate in the halo.  The enhancement of the
annihilation signal can be characterized by ratio $\eta = (I_{\rm
cl}+I_{\rm hom})/I_{\rm hom}$, where $I_{\rm cl}$ is the
annihilation signal from the clumps, and $I_{\rm hom}$ - from
homogeneously distributed DM particles with the NFW density
profile in the Galactic halo.  The main contribution to $\eta$ is
given by $\nu \simeq 2.5$ and $M \simeq M_{\rm min}$.  The signal
enhancement $\eta$ is shown numerically in
Figs.~\ref{bet1l}--\ref{bet18l}.  One can see that practically
for all allowed values of primeval perturbation spectrum index
$n_p \geq 1.0$ the annihilation signal from clumps gives the
dominant contribution.  This result does not depend on the
properties of DM particles.

The observations favor the spectrum index $n_p=1.0$ \cite{wmap}.
The enhancement of the annihilation signal for this value of
$n_p$ is described by factor 2 - 5 for different $\beta$ with
uncertainties due to values of $M_{\rm min}$ and other parameters.

The clumps which give the dominant contribution to the
annihilation signal have approximately the following properties
in the case $n_p=1$:  The mass $M \sim M_{\rm min}$ and $\nu \sim
2.5$, the radius $R \simeq 3.6 \times 10^{15}$~cm and the radius
of the core $R_c \simeq 1.8 \times 10^{14}$~cm, the mean internal
density of the clump $\bar\rho_{\text{int}}\simeq
2.5\times10^{-22}$~g~cm$^{-3}$, the fraction of the halo mass in
the form of these clumps $\xi_{\text{int}} \sim 0.002$, and the
mean number density of these clumps in the halo $n_{\rm cl} \sim
25$~ pc$^{-3}$.

Recently HEAT collaboration detected excessive flux of cosmic ray
positrons at energy $E \sim 10$~GeV \cite{heat1}.  According to
\cite{Baltz}, if this positron flux is produced by annihilation of
neutralinos the enhancement factor of order of 30 is needed. The
calculations presented here show that such enhancement can be
reached in the considered scenario in case of extreme combination
of parameters.

\begin{acknowledgments}
This work has been supported in part by the INTAS grant No.
99-1065; V.D. and Yu.E. are supported also by the RFBR grants
No.~03-02-16436-a and No. 01-02-17829.
\end{acknowledgments}

\appendix

\section{Cross sections of neutralino scattering off electrons and neutrinos}

As neutralino we shall consider here a pure bino ($\chi=\tilde
B$). The Lagrangian for interaction of bino with left and right
components of a fermion $f$ can be written (see e.~g.
\cite{jungman,haber}) as
\begin{eqnarray}
{\mathcal L}_{f\tilde f\chi}&=&-g\sqrt{2}\,\,{\rm
tan}\,\theta_{\text{W}}(e_f-T^L_{3f}) \bar f P_R\chi\tilde f_L
\nonumber \\
&&+g\sqrt{2}\,\,{\rm tan}\,\theta_{\text{W}}e_f\bar
fP_L\chi\tilde f_R, \label{lagrang}
\end{eqnarray}
where $g$ is SU(2) coupling constant, $\theta_{\text{W}}$ is the
Weinberg angle ($\sin^2\theta_{\text{W}}=0.231$), $e_f$ is
electric charge of the fermion $f$ in the units of electron
charge, $T^L_{3f}$ is the projection of weak isospin for $f_L$,
$P_R=1/2(1+\gamma_5)$ is a projection operator which cuts the
left component from the operator $\bar f$ in Eq.~(\ref{lagrang});
~  $\tilde f_L$ is the left sfermion. The first term in the
Lagrangian (\ref{lagrang}) is ${\mathcal L}_{f_L\tilde f_L\chi}$,
the second - ${\mathcal L}_{f_R\tilde f_R\chi}$. When $f$ is
ultra-relativistic in the frame where neutralino is at rest,
there is no interference for scattering of the left and right
components of the fermions (the interference terms are
proportional to $m_f$). Therefore, we shall calculate $f\chi$
cross-section for left $f_L$ and right $f_R$ fermions separately.

Scattering of the left fermion with $e_f=-1$ and $T_{3f}=-1/2$
(e.~g. $e$, $\mu$, $\tau$) off bino are described by the two
diagrams in $s$- and $u$-channels as shown below

\bigskip
\begin{picture}(100,50)
\put(5,20){\vector(1,1){5}} \put(5,40){\vector(1,-1){5}}
\put(35,30){\vector(1,1){5}} \put(35,30){\vector(1,-1){5}}
\put(67,20){\vector(1,0){10}} \put(67,40){\vector(1,0){10}}
\put(77,20){\vector(1,0){20}} \put(77,40){\vector(1,0){20}}
\put(10,25){\line(1,1){5}} \put(10,35){\line(1,-1){5}}
\multiput(15,30)(4.4,0){5}{\line(1,0){2.4}}
\put(40,25){\line(1,-1){5}} \put(40,35){\line(1,1){5}}
\multiput(87,20)(0,4.4){5}{\line(0,1){2.4}}
\put(97,20){\line(1,0){10}} \put(97,40){\line(1,0){10}}
\put(-1,40){$k_1$} \put(61,40){$k_1$} \put(47,40){$k_2$}
\put(109,20){$k_2$} \put(-1,20){$p_1$} \put(61,20){$p_1$}
\put(47,20){$p_2$} \put(109,40){$p_2$} \put(10,39){$f_L$}
\put(37,39){$f_L$} \put(77,42){$f_L$} \put(97,15){$f_L$}
\put(10,20){$\chi$} \put(39,20){$\chi$} \put(77,15){$\chi$}
\put(97,42){$\chi$} \put(25,24){$\tilde f_L$} \put(89,30){$\tilde
f_L$} \put(25,5){$s$} \put(87,5){$u$}
\end{picture}

The standard calculations for matrix elements give for
$|M|^2=|M_s|^2+|M_u|^2+2{\rm Re}(M_sM_u^*)$:
\begin{equation}
|M_s|^2=\frac{1}{2}(g\tan \theta_{\text{W}})^4
\frac{(k_1p_1)(k_2p_2)}{(s-\tilde m_L^2)^2};\label{matrms}
\end{equation}
\begin{equation}
|M_u|^2=\frac{1}{2}(g\tan \theta_{\text{W}})^4
\frac{(p_1k_2)(k_1p_2)}{(s-\tilde m_L^2)^2};\label{matrmu}
\end{equation}
\begin{equation}
 M_sM_u^*=-\frac{1}{4}(g\tan \theta_{\text{W}})^4
 \frac{m_{\chi}^2(k_1k_2)}{(s-\tilde m_L^2)(u-\tilde m_L^2)}.
 \label{matrmsmu}
\end{equation}
Cross-section for the $f_L+\chi\to f_L+\chi$ scattering at angle
$\theta_{12}$ in the system where neutralino is at rest is given
by
\begin{equation}
\left(\frac{d\sigma_{\text{el}}}{d\Omega}\right)_{f_L\chi}=
\frac{1}{64\pi^2s}|M|^2=
\frac{\alpha_{\text{e.m.}}^2}{8\cos^4\theta_{\text{W}}}
\frac{\omega^2(1+\cos\theta_{12})}{(m_{\chi}^2-\tilde m_L^2)^2},
\label{crosss1}
\end{equation}
where $\omega\gg m_f$ is energy of $f_L$ in the system where
neutralino is at rest, $m_{\chi}$ is the neutralino mass and
$\tilde m_L$ is the mass of the left sfermion.

Let us consider now $f_R+\chi\to f_R+\chi$ scattering described
by the second term in r.h.s. of Eq.~(\ref{lagrang}). The diagrams
are identical to that in the figure after substituting $f_L\to
f_R$ and $\tilde f_L\to \tilde f_R$. Since traces do not change
when $P_L\to P_R$, the expressions
(\ref{matrms})-(\ref{matrmsmu}) remain the same, changing only
due to coupling constant which increases twice (see
Eq.~(\ref{lagrang})). Therefore, we obtain
\begin{equation}
\left(\frac{d\sigma_{\text{el}}}{d\Omega}\right)_{f_R\chi}=16
\left(\frac{d\sigma_{\text{el}}}{d\Omega}\right)_{f_L\chi}.
\label{crosss2}
\end{equation}
In this paper we are interested in $\nu+\chi\to\nu+\chi$ and
$e+\chi\to e+\chi$ scattering. In the former case the
cross-section is given by Eq.~(\ref{crosss1}), and in the latter
case - by the sum of $f_L+\chi\to f_L+\chi$ and $f_R+\chi\to
f_R+\chi$ scattering, i.e. it is by factor 17 larger than the
cross-section (\ref{crosss1}).

\section{Kinetic equation}

In this Appendix we shall study the stage of kinetic equilibrium
and the stage after its breaking in the common formalism of
kinetic equation similar to \cite{gzv12} and using approach of
\cite{svvbm}. We shall confirm in this way the results of
Sec.~\ref{smmin} and clarify the difference in calculations of
$M_{\rm min}$.

Following \cite{svvbm} we introduce the neutralino distribution
function $f(x,p,t)$ over comoving coordinates $\vec x$ and
momenta $\vec p=ma^2\dot{\vec x}$  (with this definition momentum
is constant for freely moving particles).  The neutralino density
is
\begin{equation}
\rho(x,t)=\frac{m}{a^3}\int
d^3pf(x,p,t)=\bar\rho_{\chi}(t)(1+\delta(x,t)) . \label{rhobb}
\end{equation}
The kinetic equation with the collision term of the Fokker-Planck
type \cite{fk} can be written as
\begin{equation}
\frac{\partial f}{\partial t} \!+\! \frac{p_i}{ma^2}\frac{\partial
f}{\partial x_i} \!-\! m\frac{\partial \phi}{\partial
x_i}\frac{\partial f}{\partial p_i} \!=\! D_p(t)\frac{\partial
}{\partial p_i}\!\left(\frac{p_i}{mTa^2}f \!+\! \frac{\partial
f}{\partial p_i}\right)\!, \label{mainbb}
\end{equation}
where $\phi$ is the gravitational potential, which can be
neglected at the considered epoch $t\leq t_{\text eq}$, $T(t)$ is
the temperature of the ambient plasma given by Eq.~(\ref{ttime}),
and $D_p(t)$ is the diffusion coefficient in the momentum space.
According to \cite{fk}
\begin{equation}
D_p(t)=\frac{40}{3} \int d\Omega\int d\omega\, n_0(\omega) \left(
\frac{d\sigma_{{el}}}{d\Omega} \right)_{f_L\chi} (\delta p)^2.
\label{bebb}
\end{equation}
The number 40 in Eq.~(\ref{bebb}) comes from the counting of
degrees of freedom in neutralino-fermion scattering as in the
Sec.~\ref{smmin}.

The equation (\ref{mainbb}) with the diffusion coefficient
(\ref{bebb}) coincides with Eq.~(16) from \cite{gzv12} except the
numerical factor in $D_p$ which is of order of unity.

\subsection{Kinetic decoupling}

Let us consider an exit of neutralinos from the kinetic
equilibrium (decoupling) in the homogeneous universe, when
$\partial/\partial x_i$ terms in Eq.~(\ref{mainbb}) can be
neglected.  The temperature of neutralino gas $T_{\chi}$ is
defined as
\begin{equation}
\int p_ip_jfd^3p=\bar\rho_{\chi}a^5T_{\chi}(t).
\end{equation}
Multiplying Eq.~(\ref{mainbb}) by $p_ip_j$ and integrating it
over $d^3p$ one obtains
\begin{equation}
\frac{dT_{\chi}}{dt}+2\frac{\dot a}{a} T_{\chi}
-\frac{2D_p(t)}{ma^2}\left( 1-\frac{T_{\chi}(t)}{T(t)} \right)=0,
\label{ttbb}
\end{equation}
The initial condition for Eq.~(\ref{ttbb}) can be chosen at the
moment of freezing $t=t_f$ as in \cite{gzv12}, or more
conveniently at any $t_i$ from interval $t_f< t_i\ll t_d$, as
$T_{\chi}(t_i)=T(t_i)$, where $T$ is the temperature of ambient
plasma.  Solution of Eq.~(\ref{ttbb}) (see below) gives
transition of ratio $r(t)=T_{\chi}(t)/T(t)$ from $r=1$ to $r_d<1$
within some time interval, determined by $r_d$.  Any value of $t$
in this interval can be taken as definition of decoupling time
$t_d$.  Eq.~(\ref{ttbb}) and its solution can be simplified using
the dimensionless time $\tau= t/t_d$. Characteristic time $t_d$
is naturally emerged from dimension parameters entering the
diffusion coefficient, and up to numerical coefficient it
coincides with $t_d$ determined in Sec. \ref{smmin}.  The
transition time interval fixes this numerical coefficient with
some uncertainty, and we obtain indeed $t_d$ (and hence
$T_d=T(t_d)$) approximately equal to those given by
Eqs.~(\ref{tsmd}) and (\ref{tbigd}) in Sec.~\ref{smmin}.  The
solution of Eq.~(\ref{ttbb}) in terms of $\tau=t/t_d$ is given by
\begin{equation}
 \frac{T_{\chi}(t)}{T_d}\!=\!\frac{1}{\tau}\!\!\left(\!
 \tau_i^{-1/2}e^{1/4\tau^2-1/4\tau_i^2}\!+\!
 \frac{1}{2}e^{1/4\tau^2}\!\!\int\limits_{\tau_i}^{\tau}
 \!\!d^3xx^{-5/2}e^{1/4x^2} \!\right)\!. \label{soltau}
\end{equation}
The asymptotic forms of solution (\ref{soltau}) are given by
$T_{\chi}/T_d=\tau^{-1/2}$ for $\tau\ll1$ and
$T_{\chi}/T_d=\tau^{-1}\Gamma(3/4)/2^{1/2}$ for $\tau\gg1$ as it
must be. From solution (\ref{soltau}) it is seen that transition
from kinetic equilibrium of neutralino with relativistic fermions
to the non-equilibrium regime proceeds very fast.  By this reason
our consideration of diffusion and free streaming independently
in Sec.~\ref{smmin} is well justified.

\subsection{Diffusion}

Consider Eq.~(\ref{mainbb}) before kinetic decoupling, $t\ll
t_d$.  One can find the first two moments by integrating first
time Eq.~(\ref{mainbb}) over $d^3p$ and second time over
$p_id^3p$.  Inserting the first of the obtained equation into the
second one we obtain the following equation for the Fourier
components:
\begin{equation}
\frac{\partial^2 \delta}{\partial^2 t}+ 2\frac{\dot
a}{a}\frac{\partial \delta}{\partial t}
+D_p(t)\frac{1}{mTa^2}\frac{\partial \delta}{\partial t}=
\frac{k_ik_j}{\bar\rho_{\chi}a^7m} \int p_ip_jfd^3p.
\label{main1}
\end{equation}
The r.h.s.  of the (\ref{main1}) has a tensor form
\begin{equation}
\frac{1}{\bar\rho_{\chi}a^7m}\int
p_ip_jfd^3p=E\delta_{ij}+'Fk_ik_j,
\end{equation}
where the isotropic part $E=T_{\chi}\delta_k/a^2m$ for any
$\tau$, while $F$ depends on time $t$.  In the limit $\tau\ll 1$
we may put $F=0$ and neglect the first and second terms in
(\ref{main1}).  The resultant equation coincides with diffusion
equation (\ref{diff}) with the same diffusion coefficient
(\ref{ddd}) and has the same solution.

In \cite{gzv12} only this diffusion limit of the general kinetic
equation (\ref{mainbb}) has been considered.

\subsection{Free streaming}

In the limiting case $\tau\gg1$, i.e. after decoupling, the
(\ref{mainbb}) has a simple form
\begin{equation}
\frac{\partial f}{\partial t}+ \frac{p_i}{ma^2}\frac{\partial
f}{\partial x_i}=0, \label{main2}
\end{equation}
with the solution
\begin{equation}
f\propto \exp\left[\frac{ik_jp_j}{ma_d}g(t)\right], \label{bnv}
\end{equation}
where $g(t)$ is the same function as (\ref{defg}). The solution
(\ref{bnv}) is valid with a good accuracy also at $\tau\geq1$,
because according to (\ref{soltau}), kinetic decoupling proceeds
very fast. Integrating (\ref{bnv}) over $d^3p$ with initial
condition
\begin{equation}
f(t_d)= (2\pi T_dma_d^2)^{-3/2}
\exp\left\{-\frac{p^2}{2T_dma_d^2}\right\},
\end{equation}
one obtains
\begin{equation}
n_{\vec k}(t) = n_{\vec k}(t_d) e^{-(1/2)k^2g^2(t)T_d/m_{\chi}},
\end{equation}
and then Eqs.~(\ref{lambdafs}), (\ref{mfs}) and (\ref{mminnum})
from Sec.~\ref{smmin}.

\end{document}